\documentclass{article}
\usepackage{graphicx} 
\usepackage{authblk}
\usepackage{geometry}
\usepackage{xcolor}
\usepackage{amsmath}
\usepackage{array}
\usepackage{multirow}
\usepackage{amsfonts}
\usepackage{subcaption}
\usepackage{listings}
\usepackage{xcolor}
\usepackage[normalem]{ulem}
\usepackage{siunitx}
\usepackage{caption}

\lstdefinelanguage{Stan}{
  keywords={data, int, real, parameters, model, generated, quantities, for},
  sensitive=true,
  comment=[l]{//},
  morecomment=[s]{/*}{*/},
  morestring=[b]",
}

\definecolor{red}{named}{black} 

\title{Bayesian Modeling for Aggregated Relational Data: A Unified Perspective}
\date{\today}

\usepackage[style=authoryear,natbib, uniquename=false, maxcitenames=2]{biblatex}
\author[1]{Owen G. Ward}
\author[2]{Anna L. Smith}
\author[3]{Tian Zheng}

\affil[1]{Department of Statistics and Actuarial Science, Simon 
Fraser University}
\affil[2]{Dr. Bing Zhang Department of Statistics, University of Kentucky}
\affil[3]{Department of Statistics, Columbia University}
\affil[ ]{owen\_ward@sfu.ca, Anna.L.Smith@uky.edu, tian.zheng@columbia.edu}

\begin{document}
\maketitle

\begin{abstract}
Aggregated relational data is widely collected to study social networks,
in fields such as sociology, public health and economics.
Many of the successes of ARD inference have been driven by increasingly complex Bayesian models,
which 
provide principled and flexible ways of reflecting dependence patterns and biases encountered 
in real 
data.
{\color{red}In this work we} provide researchers with a 
{\color{red}unified}
collection of Bayesian implementations of existing models for ARD, within the state-of-the-art Bayesian sampling language Stan.
Our implementations incorporate within-iteration rescaling procedures by default, 
improving algorithm run time and convergence diagnostics. 
{\color{red}Estimating ARD parameters requires carefully balancing model complexity against computational cost and data requirements, yet this trade-off has received relatively limited systematic attention in the literature.
Moreover, general model comparison tools applicable across a wide range of ARD models remain underdeveloped, and existing approaches often require substantial expertise in Bayesian computation and software.}
Using synthetic data, we demonstrate how {\color{red} well} competing models recover true personal network sizes and subpopulation sizes and how existing posterior predictive checks compare across a range of Bayesian ARD models. 
We provide code to leverage Stan's modeling framework for {\color{red}exact $K$-fold} cross-validation, and {\color{red} explain why approximate leave-one-out estimates often fail for many ARD models}. 
{\color{red}This work highlights important connections and future directions
in Bayesian modeling of ARD, providing practical guidance for selecting and evaluating Bayesian ARD
models.}

\end{abstract}


\section{Introduction}

Aggregated relational data (ARD) has been widely studied in the statistical literature
in recent years, and is used 
in solving a variety of key problems in the study of social networks.
These include estimating the size of hidden populations \citep{bernard2010counting}, 
estimating personal network sizes \citep{mccormick_how_2010},
understanding global network structures \citep{breza2023consistently} and fitting complex latent variable models to
massive network data \citep{jones2021scalable}. 
ARD has been used to recover network structure, {\color{red}using}
data which is less expensive 
and more easily collected than traditional network data \citep{breza2020using}.
Many of the successes of ARD models have been driven by the {\color{red}use}
of Bayesian implementations, providing a principled and flexible way to fit and interpret 
these
models for real data. As such, an understanding of Bayesian modeling for ARD is key
to realize the potential of these statistical tools.
These Bayesian implementations also provide
practitioners with a wealth of procedures to
examine the fit of these models,
{\color{red}which have not been systematically studied or presented in a unified computational framework.}

In this paper we 
{\color{red}present a unified and practical}
overview of statistical methods for ARD and highlight
the benefits of utilising Bayesian {\color{black}inference} for such models. 
{\color{red}We begin by outlining}
the key features of Bayesian modeling for ARD, 
before {\color{red}presenting}
a systematic collection of Bayesian
implementations of models for ARD. 
We describe the key properties that each of these 
models 
{\color{red}are designed to capture,}
while contrasting the computational challenges required to fit more complex
models.
We carefully describe connections and differences between model specifications and highlight salient
assumptions and data requirements for each model.
We demonstrate how Bayesian modeling practices can resolve many of the issues which occur when fitting 
existing ARD models, through worked examples with popular modern ARD models for realistic simulated 
data.
We {\color{red}then} describe how these implementations 
provide immediate access to existing Bayesian model checking techniques {\color{black}
which can be used for initial assessment of competing Bayesian ARD models.}
{\color{black}Finally, we provide a discussion of}
Bayesian model checking procedures specific to
ARD and {\color{black}identify important considerations and challenges in this context.}


{\color{red}Using illustrative examples, we narrate the practical considerations when choosing between Bayesian ARD models,
which has not been systematically addressed previously in the literature.
We synthesize existing approaches while explicitly highlighting their assumptions and advantages, along with practical trade-offs and considerations.}
We {\color{red}include}
documented code for each model and model checking procedure we consider here,
along with simulated data which can be used to investigate these models directly.
{\color{black}This provides researchers with a ready to use suite of implementations of
Bayesian ARD models and tools for model checking and comparison.}
{\color{red}While previous work has summarized existing ARD models, we explicitly
describe the relevant
practical challenges that arise when fitting and working
with these models.
As such, this work serves as both a reference and a practical 
guide for statisticians and practitioners working with ARD data.}

\section{Aggregated Relational Data} \label{sec:data}

Before introducing Bayesian models for ARD we first provide a brief review of
the notation for data modeled using ARD, along with an overview of 
the common tasks of interest with such data.
We note that existing work such as \citet{laga2021thirty, mccormick2020network}
provides a comprehensive review of methods for the specific task of estimating
subpopulation sizes using ARD.
Here we instead focus on the practical concerns that arise with fitting
models for ARD, which we highlight using a selection of popular models in the literature.
{\color{red}These practical considerations have not previously been systematically documented.}

\subsection{Notation}

Aggregated relational data is used to collect information
about actors in a social network. Suppose we have a population 
of size $N$. We can then record the presence or absence 
of a link between all nodes in this network using an
adjacency matrix $A= \left[A_{ij} \right]_{N\times N}$.
Here $A_{ij}=1$ if person $i$ ``knows" person $j$.
If this matrix $A$ was observed then we would 
be able to compute the degree $d_i$ of any actor in this 
network, where $d_i=\sum_{j}A_{ij}.$ In practice,
obtaining $A$ is not feasible.

If estimating the degree of an individual is the main 
goal then this could be achieved by asking them 
if they know $m$ randomly chosen members of the population \citep{mccormick_how_2010}. 
However this is inefficient and requires a large $m$
to give good estimates.
Originally, ARD arose in the context of estimating the size of
hard to reach populations in a network where this underlying $A$ is not observed 
\citep{bernard1991estimating}. 

The widely used approach to obtain information 
in this setting, when $A$ is not observed,
is to instead ask a sample of size $n$ about their relationship
to several subpopulations in the network.
These are commonly questions of the form ``How many X's do you know?",
where X is a specific subpopulation within the population. 
For a node $i$, we record the number of people they know in 
subpopulation $k$ and represent this as $y_{ik}$. 
We will denote the set of nodes in subpopulation $k$ in the population
as $G_k$.
If this is done for $K$
subpopulations then this results in the $n\times K$ ARD matrix $Y=[y_{ik}]_{n\times K}$
where 
$$
y_{ik} =\sum_{j \in G_k}A_{ij}.
$$

ARD of this form can then readily be used to estimate individual 
node degree, such as the scale-up estimator of \citet{killworth1998estimation}
if the size of these subpopulations in the population, $N_k$,
is known (such as through external data sources, including the U.S. Census).
This data can also be used to estimate the 
size of one or more hidden subpopulations, given these degree estimates 
\citep{feehan2016generalizing, maltiel2015estimating,kunke2024comparing}.
This formulation of $y_{ik}$ as a sum of Bernoulli random events
is useful as it allows for this quantity to be modeled (either exactly or approximately)
as Poisson distributed, where $y_{ik}\sim \text{Poisson}(\lambda_{ik})$
and $\lambda_{ik}=\sum_{j \in G_k}P(A_{ij}=1)$.
This is the most commonly used formulation of statistical models for ARD 
and the models we consider below will work from this initial
starting point.

{\color{black}\subsection{Model-based exploration of ARD characteristics}}
{\color{black}
Different applications may require a chosen model to capture different social characteristics and it is 
{\color{red}essential}
that a user is able to identify which social phenomena are explicitly incorporated within a given model.
In this section we briefly review some of the key properties of popular ARD models and how to {\color{red}assess} whether a fitted model appropriately accounts for social structures exhibited in the observed data, before then using the estimated model for inference.
Examining each of these properties for a fitted model can help guide how to expand that model and better describe the underlying social dynamics.
Figure \ref{fig:model_diagram} visually represents an ordering of ARD models by the complexity of the social dynamics explicitly incorporated through model parameters and structure.
This ordering can be used to guide the iterative model checking and expansion procedure that underlies the flexibility and power of Bayesian methods \citep{gelman2020bayesian}.
Finally, we 
{\color{red}summarize}
methods to fit ARD models and highlight how advanced Bayesian methods can accommodate large scale ARD and complex structures.}
\\
\paragraph{Estimating degrees}%
A key task for ARD models is to estimate the degree
of nodes in the ARD sample
(also known as personal network size),
$d_1,\ldots, d_n$. 
By using synthetic data we will be able to 
compare the estimated node degrees under
each of the models we fit to a ground truth, providing 
a tool for model comparison and evaluation. While this model
evaluation is not available in real data, we will use it to 
verify the results obtained from other model checking methods.
For observed data, examining histograms or kernel density estimates of the estimated degrees, as we do in Figure \ref{fig:latent_true_degree}, can help verify whether the fitted model provides reasonable estimates for the particular scientific context under study.  \textcolor{black}{For example, if the degree distribution is not well estimated under the \citet{handcock2004likelihood} model, further expanding the model by accounting for overdispersion may improve overall model fit and degree estimation (see Figure \ref{fig:model_diagram} for an ordering of Bayesian ARD models by complexity).  Similarly, if there is large variability in degrees across ego groups, expanding the model to explicitly incorporate ego demographics (e.g., a model from the right column of Figure \ref{fig:model_diagram}) may prove useful}.  

\paragraph{Recovering known subpopulation sizes}  ARD collection includes reference subpopulations for which the subpopulation size is known.  Reference groups are commonly implemented via names (e.g., how many people named ``Christina'' do you know?), since the subpopulation size can be estimated from government records (e.g., the Social Security Administration).   The Bayesian ARD models we study below treat the reference subpopulation sizes as random variables (model parameters) and the count obtained from a government record or similar database as a single observation of this variable.  Our Bayesian models naturally provide posterior distributions for the size of these reference distributions.  Disagreement between this posterior distribution and the known subpopulation sizes,
{\color{black}potentially after applying a rescaling method
to obtain interpretable size estimates,}
is another tool for diagnosing poor model fit.
{\color{black} Evaluating ARD models in terms of estimation 
of known subpopulations has been considered as a form of leave-one-out estimation
\citep{laga2023correlated}, numerically quantifying the performance of a model.} \textcolor{black}{As will be discussed in Section \ref{sec-iden}, subpopulation size parameters are typically confounded with personal degree or gregariousness parameters, which is why a rescaling procedure is typically required.  The accuracy of subpopulation size recovery for reference groups can be used to compare different rescaling techniques, different prior distributions, and different model structures, as we demonstrate with synthetic data in Section \ref{sec-checking}.}

\paragraph{Nonrandom mixing}
While sociologists' definitions of ``social structure'' vary and are often quite nuanced, 
one particular definition is useful for the statistical modeling of ARD: ``social 
structure is the difference in affiliation patterns from what would be observed if people 
formed friendships entirely randomly'' \citep{blau_1974}.  Under this definition, 
quantifying differences in the likelihood of knowing different subpopulations, and how 
those may vary across different types of individuals (e.g., as a function of basic 
demographics like gender or age), is of critical scientific interest in the study of 
social relationships.  This concept is especially easy to study within so-called 
overdispersed models, where we explicitly allow the likelihood of ties between an 
individual and a subpopulation to vary, {\color{black}and having 
this additional flexibility is common in most advanced ARD models.}
This is closely related to the ``barrier effects'' problem that has been identified for 
scale-up estimators, which leverage ARD to estimate the size of hard-to-reach 
populations, such as the unhoused or people who misuse drugs 
\citep{killworth_et_al_2006}.  Barrier effects, also called nonrandom mixing, are said to 
occur ``whenever some individuals systematically know more (or fewer) members of a 
specific subpopulation than would be expected under random mixing'' 
\citep{mccormick_how_2010}.
In Section \ref{sec:checking}, we consider two models which account for nonrandom mixing and demonstrate how model fit metrics can help evaluate whether an ARD model that {\color{black}captures such} mixing provides a better explanation of the data under study. {\color{black}Being able to identify this
for a given problem is an important tool.}  \textcolor{black}{In Figure \ref{fig:model_diagram} and Section \ref{mod-more}, we delineate four increasingly complex model structures that reflect nonrandom mixing.  Degree estimation and subpopulation recovery can be compared across these models to determine which version of nonrandom mixing structure provides the best model fit.  As is typical in the Bayesian workflow \citep{gelman2020bayesian}, we may begin with a simpler model and continue iteratively adjusting and expanding the model to incorporate more complex structures as dictated by evidence for lack of fit.}

\paragraph{Fitting and Evaluating these Models}
Regardless of the model properties considered
above,
an important practical consideration with any 
ARD model is the 
computational 
tools required to fit these models along with how to assess the
validity of the estimated fitted model.
A variety of Bayesian approaches have been widely used, from straightforward Gibbs-Metropolis algorithms to flexible probabilistic programming languages to advanced variational inference, and provide potential
tools for applications of many different scales.
With regards to fitting large complex 
models
quickly, variational inference \citep{blei2017variational}
is a powerful Bayesian inference approach for such tasks,
and has already been used in the context of large scale ARD \citep{jones2021scalable}.
Modern software for Bayesian inference also provides a large suite
of ready to use tools to assess model convergence and diagnose problems with
general Bayesian estimation,
which can be considered for ARD models also \citep{gelman2013bayesian, gelman2020bayesian}. After introducing several
Bayesian ARD models below we will describe how these tools can be used 
and extended when comparing ARD models.
{\color{red}Despite the availability of these computational tools,
their application and adaptation to ARD models has not been systematically studied, particularly in the context of comparing competing model specifications and performing model selection.}

\section{Bayesian Models for ARD}
\label{sec:models}
\begin{figure}[ht!]
	\centering
   \includegraphics[width=\textwidth]{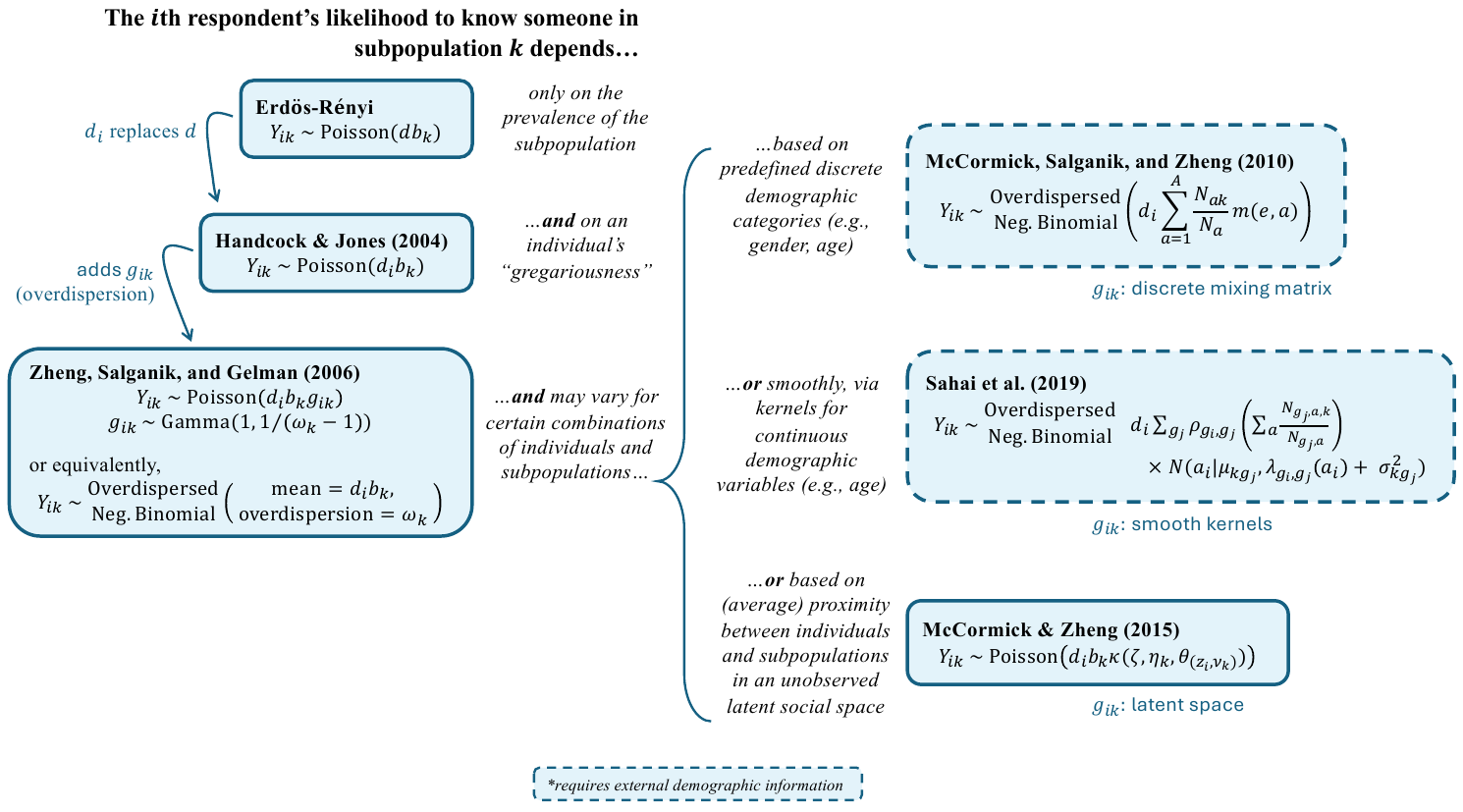}
	\caption{Links between Bayesian models for aggregated relational data. All models which assume the overdispersed negative binomial distribution also include an overdispersion parameter, $\omega_k$.}
	\label{fig:model_diagram}
\end{figure}

A wide variety of Bayesian ARD models have been proposed in the past two decades,
giving practitioners many potential models to consider. These models have become
increasingly sophisticated, in an attempt to represent the complex dependence structure
inherent in social network data. In this section we present a selective overview of
some of the major contributions to Bayesian modeling of ARD. We start with simple
null models initially in Section~\ref{mod-null}, highlighting the connection
to classical network models and how these models are limited in the variation they can capture.
We then consider how these simple models have been extended to
capture more appropriate social structure via nonrandom mixing 
in Section~\ref{mod-more}. We consider models which can also incorporate
covariate information, in the form of demographic knowledge, in Section~\ref{mod-cov}.
In Section~\ref{mod-nsum} we briefly discuss the important class 
of ARD models related to the network scale-up procedure, a specific form of ARD model.
Table \ref{tab:models} and Figure \ref{fig:model_diagram} visually represent the connections and similarities across the different model specifications we discuss in this section.
We will then compare these models in the context of simulated data in subsequent sections, 
examining both the social structure that they can recover along with performing
model checking and comparison.
{\color{red}While many of these models have been built on previous (simpler) models, their 
comparative performance and associated computational tradeoffs have not been systematically examined, which we address in this work.}

{\color{black}\subsection{Null models}\label{mod-null}}

Before discussing the advances made in Bayesian modeling for ARD, we begin with two simple data models, 
which will be used to
demonstrate model fitting and how these {\color{black}initial}
models fail when fit to data: the 
Erd\"os-R\'enyi model and what we will call the null model, following 
\citet{zheng_how_2006}'s description.  
The Erd\"os-R\'enyi model is a classical mathematical model for graphs 
which 
assumes that (acquaintanceship) ties are formed completely at
random \citep{erdos1960evolution}.  
Under this assumption, the analogous model for ARD is 

$$
y_{ik} \sim \text{Poisson}(db_k),
$$

where $d_i = \sum_{j=1}^N P(\text{person $i$ is acquainted 
with person $j$})$, 
the expected degree of individual $i$, is assumed constant 
under this model 
(i.e., $d_i = d \>\forall i$) and $b_k$ allows the likelihood of 
forming ties 
to vary by subpopulation $k$.
The Erd\"os-R\'enyi model assumes that ties are formed purely at random, which implies that $b_k \approx N_k/N$, the prevalence of
subpopulation $k$ in the population.  
Of course, in practice we also observe heterogeneity in the number of people each survey respondent reports being acquainted with, which would result in super-Poisson variation of the $y_{ik}$.

In the null model, which follows work by \citet{handcock2004likelihood}, 
we allow the propensity to form connections to differ across different 
survey respondents and different subpopulations, {\color{black}
driven by individual degree parameters $d_i$ with}

$$
y_{ik} \sim \text{Poisson}( d_i b_k ). 
$$

Notably, this does not allow for an interaction effect, where we might 
observe different propensities across subgroups
{\color{black}as there is no term which depends on both
the node $i$ and the subpopulation $k$.}
Using the example in 
\citet{zheng_how_2006}, the probability that a reader of this journal 
knows someone who is in prison is likely different than the probability 
that an individual without a high school degree knows someone who is in prison, {\color{black}but that is not possible in this model, if they have the same degree. 
An interaction effect is required to introduce this additional flexibility.}

{\color{black}\subsection{Models with nonrandom mixing}\label{mod-more}}

\paragraph{Overdispersed model}
\label{sec:2006_model}
As indicated by the example referenced above, \citet{zheng_how_2006} observed that individuals often displayed great variation in their propensity to form ties to some groups (e.g., males in prison), but not to others (e.g., twins, people named Michael or Nicole).  In other words, ties associated with some subgroups or individuals demonstrated extra variability not accounted for by the null model, i.e., overdispersion.  \citet{zheng_how_2006}'s model treats this additional variability as its own source of information and models it directly via a dedicated overdispersion parameter.  The overdispersed model assumes a log-linear form with

$$
y_{ik} \sim \text{Poisson} \left( \exp \left\{ \alpha_i + \beta_k + \gamma_{ik} \right\} \right),
$$ 
or, letting $d_i = e^{\alpha_i}$, $b_k = e^{\beta_k}$, and $g_{ik} = e^{\gamma_{ik}}$, 
equivalently,
\begin{align} \begin{split} \label{model_overdispersed}
    y_{ik} &\sim \text{Poisson}( d_i b_k g_{ik}), \\
    g_{ik} &\sim \text{Gamma} \left( {\color{black}\text{mean}=\ } 1, {\color{black}\text{shape} =\ } \frac{1}{\omega_k - 1} \right). 
\end{split} \end{align}

Importantly, the model does not attempt to estimate all of the individual $g_{ik}$'s, 
but only certain properties of their distribution. 
Further, the distributional assumption for $g_{ik}$ allows us to easily
integrate out the $\gamma$'s, yielding the following simplified model specification,

\begin{align}
    y_{ik} \sim \text{Neg. Binomial}\Big( & \mu_{ik} = d_i b_k,  \\
    & \text{ overdispersion = } \omega_k \Big). \notag
\end{align}

The overdispersion parameter, $\omega_k$, can be interpreted as (1) an up-scaling factor for the variance, since $Var(y_{ik}) = \omega_k E[y_{ik}]$ and (2) as a factor that decreases the frequency of individuals who know exactly one person in a given subpopulation relative to the frequency of those who know none, since you can show that $P(y_{ik}=1) = \frac{1}{\omega_k} P(y_{ik}=0) E[y_{ik}].$  When $\omega_k = 1$ (which corresponds to Var[$g_{ik}$]=0), the overdispersed model reduces to the null model.
{\color{red}In \citet{zheng_how_2006}}
the $b_k$'s are estimated freely, but renormalized within the Markov Chain Monte Carlo (MCMC) sampling algorithm so that $b_k \approx N_k/N$ for the rarest names.

\paragraph{Latent Space model}
\label{model-latent}
\citet{mccormick_latent_2015} assume that individuals and subpopulations are positioned in a latent social space, represented as the surface of a $p$-dimensional hypersphere, $\mathcal{S}^{p+1}$. Survey respondents and subpopulation members who are closer together in this latent social space are more likely to be tied; in this sense, latent distances represent additional variability due to social structure that is unaccounted for in the null model.  \citet{mccormick_latent_2015} begin with a log-linear model for the completely observed graph and arrive at the following specification: 
\begin{align}
    y_{ik} &\sim \text{Poisson}( d_i b_k \kappa(\zeta, \eta_k, \theta_{(z_i,\nu_k)}) ),\\
    \text{where } \kappa(\zeta, \eta_k, \theta_{(z_i,\nu_k)}) &= \frac{C_{p+1}(\zeta) C_{p+1}(\eta_k)}{C_{p+1}(0)C_{p+1}(\sqrt{\zeta^2 + \eta_k^2 + 2 \zeta \eta_k \text{cos}(\theta_{(z_i,\nu_k)})})}.
\end{align}
{\color{red}Here} $C_{p+1}$ is the normalization constant for the von Mises-Fisher distribution, which is the distribution for all the latent positions on the sphere, $\zeta$ scales the influence of the latent distances, $\eta_k$ controls the concentration of latent positions for individuals in subpopulation $k$, and $\theta_{(z_i,\nu_k)}$ is the angular distance between the latent position for individual $i$ and subpopulation $k$.  

Similar to the $g_{ik}$ parameters in the overdispersed model, \citet{mccormick_latent_2015} do not attempt to estimate latent positions for all survey respondents (egos) and subpopulation members (alters), but leverage certain properties of their distribution.  Since the alters are not actually observed in the ARD setting, \citet{mccormick_latent_2015} introduce a distribution of alters in the latent space and take an expectation over all individuals' positions in each subpopulation.  In this sense, the latent space ARD model is conditioned on an individual's \textit{expected} distance to any member of a given subpopulation.  This extends the classic latent space network model for non-aggregated social network data developed by \citet{hoff_raftery_handcock_2002} which is conditioned simply on the latent distance between individuals.

\begin{table}[ht!]
    \centering
    \begin{tabular}{ l c c m{2.5in} }
         \hline
         Paper & Expected Number of Ties & Distr. & Additional Model Parameters \\
         \hline \hline
         
         \multirow{2}{1in}{Erd\"os-R\'enyi} & $db_k$ & P & $b_k = N_k / N$ \\
         & & & prevalence of subpopulation $k$ in the population \\
         \hline
         
         \multirow{2}{1in}{\citet{handcock2004likelihood}} & $d_ib_k$ & P & $d_i = \sum_{j=1}^N p_{ij}$ \\
         &&& expected degree for individual $i$, \\
         &&& $p_{ij} = P(\text{individuals } i,j \text{ know each other})$ \\
         \hline
         
         \multirow{2}{1in}{\citet{zheng_how_2006}} & $d_ib_kg_{ik}$ & P/ONB & $g_{ik} = \lambda_{ik}/d_ib_k$\\
         &&& individual $i$'s relative propensity to know an individual in subpopulation $k$ \\
         \hline
         
         \multirow{2}{1in}{\citet{mccormick_how_2010}} &  $d_i \sum_{a=1}^A \frac{N_{ak}}{N_a} m(e,a)$ & ONB & $N_{ak} / N_a$ \\
         &&& proportion of individuals in demographic group $a$ also in subpopulation $k$\\[.15in]
         &&& $m(e,a)$ \\
         &&& mixing propensity for individual $i$'s demographic group, $e$, to know demographic group $a$ \\
         \hline
         
         \multirow{2}{1in}{\citet{sahai2019estimating}} &  $d_i \sum_{g_j} \rho_{g_i,g_j} \left( \sum_a \frac{N_{g_j,a,k}}{N_{g_j,a}} \right)$  & ONB & $\rho_{g_i,g_j}$ \\
         & $\>\>\times N(a_i| \mu_{kg_j}, \lambda_{g_i,g_j}(a_i) + \sigma^2_{kg_j})$ && mixing propensity for individual $i$'s gender group, $g_i$, to know gender group $g_j$ \\[.15in]
         &&& $N_{g_j,a,k} / N_{g_j,a}$ \\
         &&& proportion of individuals of age group $a$ and gender $g_j$ also in subpopulation $k$ \\[.15in]
         &&& $\mu_{kg_j}, \lambda_{g_i,g_j}(a_i), \sigma^2_{kg_j}$ \\
         &&& parameters of the smooth kernel that represents the propensity of an individual with gender $g_i$ and age $a_i$ to form ties to an individual of gender $g_j$ \\
         \hline 
         
         \multirow{2}{1in}{\citet{mccormick_latent_2015}} & $d_ib_k \times$ & P & $\kappa(\zeta,\eta_k, \theta_{(z_i,\nu_k)})$\\
         &  $\kappa(\zeta,\eta_k, \theta_{(z_i,\nu_k)})$ && describes how the latent space influences the probability of forming an ARD tie \\[.15in]
         &&& $\theta_{(z_i,\nu_k)}$: angular distance between individual $i$ and the center of subpopulation $k$ in the latent space \\[.15in]
         &&& $\zeta$: scales the influence of the latent positions \\
         &&& $\eta_k > 0$: concentration of alters' latent positions in subpopulation $k$ \\
         \hline
         
    \end{tabular}
    \caption{Table of model specifications and parameters for Bayesian ARD models. Each model assumes that the number of ties between individual $i$ and subpopulation $k$ is distributed according to either a Poisson (P) or overdispersed Negative Binomial (ONB) distribution.  All models that assume the overdispersed Negative Binomial distribution include an overdispersion parameter, $\omega_{k}$. In \citet{zheng_how_2006}, the Poisson model is equivalent to an overdispersed Negative Binomial model since they assume that $g_{ik} \sim \text{Gamma}(1/(\omega_k - 1),1/(\omega_k - 1))$.}
    \label{tab:models}
\end{table}


{\color{black}\subsection{Covariate-based models of nonrandom mixing}\label{mod-cov}}
A well-known principle of the social dynamics that drive network relationships is homophily, or the phenomenon in which individuals who are more similar are more likely to be connected \citep{coleman_1958,lazarsfeld_merton_1954}.  In practice, this is typically formulated in statistical models via individual-level covariates, such as gender, age, socioeconomic status, etc.  In Bayesian models for ARD
which attempt to account for this principle,
patterns of ties across similar (and dissimilar) individuals are explicitly used to account for patterns of nonrandom mixing (i.e., overdispersion).  In the ARD setting, collecting this additional information from survey respondents is relatively easy.  However, by design, we do not collect data from members of the subpopulations of interest; instead, the models that follow use external data sources to calculate the relative frequencies of each demographic variable within each subpopulation.  For example, data from the U.S. Social Security administration (SSA) can be used to construct age profiles for names.  These models do not enforce homophily explicitly (i.e., there is no assumption of {\color{red}homophily} in the model specification), but instead specify flexible parameterizations that allow for variability in the likelihood of ties based on these demographic groups and allow us to examine whether or not homophily is exhibited in the ARD we collected.

The model specification discussed below also {\color{black}allows} us to examine broader patterns of nonrandom mixing not directly implied by homophily.  Borrowing the example discussed in \citet{zheng_how_2006}, ``people with certain diseases may not necessarily associate with each other, but they could have a higher propensity to know health care workers.''
As these models require the incorporation
of covariate or demographic information, we do not consider them
for our simulation studies. However, these models provide
important extensions if such covariate information is available.
{\color{black}This is a particularly active area of research in ARD modeling, with 
recent extensions and applications including \citet{baum2023uses, quaye2023application, laga2023correlated}. Here we briefly describe two 
extensions, which allow increasingly flexible covariate information.}
\\
\paragraph{Discrete nonrandom mixing model}
\citet{mccormick_how_2010} suppose $A$ alter groups and $E$ ego groups (e.g., based on demographics such as gender or age).  
For respondent $i$ belonging to ego group $e$, 
\begin{align}
    y_{ik} \sim \text{Neg. Binomial}\Big( &  \mu_{ike} = d_i \sum_{a=1}^A \frac{N_{ak}}{N_a} m(e,a), \\
    & \text{ overdispersion = } \omega_k \Big), \notag
\end{align}
{\color{black}where $N_a$ is the (known) size of alter group $a$}, $N_{ak}$ is the (known) 
number of individuals in both subpopulation $k$ and alter group $a$, and 
$m(e,a)$ is a model parameter representing the likelihood that individuals from 
ego group $e$ know individuals from alter group $a$. In the presence of truly 
random mixing, $m(e,a) \approx \frac{N_a}{N}$.  In other words, the likelihood 
that an individual from ego group $e$ knows individuals from alter group $a$ 
will be based only on the relative size of the alter group.  In this case, 
$\mu_{ike} = d_i N_k/N \approx d_i b_k$, matching the 
overdispersed model in (\ref{model_overdispersed}).

\paragraph{Kernel smoothing model} \citet{sahai2019estimating} extend the mixing matrix based on discrete demographic categories 
of \citet{mccormick_how_2010}
to the continuous setting by specifying smooth kernels that describe how the likelihood of a tie 
depends on a continuous demographic factor, such as age.  The model in \citet{sahai2019estimating} 
specifies a smooth kernel for participant age {\color{black}$a_i$} with a multiplicative factor for 
gender {\color{black}$g_i$}, 
but could be extended to incorporate other covariates or more kernels:
\begin{align}
    y_{ik} \sim \text{Neg. Binomial}\Bigg( &  \mu_{ik} = d_i \sum_{g_j} \rho_{g_i,g_j} \left( \sum_a \frac{N_{g_j,a,k}}{N_{g_j,a}} \right)  N(a_i| \mu_{kg_j}, \lambda_{g_i,g_j}(a_i) + \sigma^2_{kg_j}), \\
    & \text{ overdispersion = } \omega_k \Bigg). \notag
\end{align}

{\color{black}Here $\lambda_{g_i,g_j}(a_i)$ is a latent factor which can be thought 
of as the bandwidth of the age mixing kernel for a given ego age $a_i$. 
$N_{k,a,g_j}$ is the number of people in subpopulation $k$ with age $a$ and gender $g_j$.
$\lambda_{g_i,g_j}(a_i)$ is parameterized
as a fourth order spline. Similarly, $\rho_{g_i,g_j}$ is the corresponding entry of a gender 
only mixing matrix. \citet{sahai2019estimating} arrive at this representation 
after approximating the alter $g_j$ group distribution using available information.
Notably, we believe this was the first ARD model to be implemented in Stan.}

{\color{black}\subsection{Bayesian Network Scale-Up Models}\label{mod-nsum}}
{\color{black}While we have discussed ARD models in general above, a large
class of ARD models have been developed extending the initial network scale 
up method (NSUM) of \citet{bernard1991estimating}.}
\textcolor{black}
{\citet{maltiel2015estimating} propose a series of models that extend the Binomial-likelihood, traditionally used in NSUM methods, to account for common patterns in ARD data. 
These models include analogs of \citet{handcock2004likelihood}'s null model (called the random degree model), of \citet{zheng_how_2006}'s overdispersed model (called the barrier effects model), and additional models that account for transmission and recall bias in the survey responses.
In Section \ref{sec:sim_data}, we will simulate synthetic ARD from \citet{maltiel2015estimating}'s barrier effects model, which is given by
\begin{align*}
    y_{ik} &\sim \text{Binomial}(d_i, q_{ik}),\\
    q_{ik} &\sim \text{Beta}(m_k,\rho_k),
\end{align*} 
where the distribution of $q_{ik}$ uses the nonstandard parameterization of the Beta distribution \citep{skellam1948probability} and a hyperprior is specified for $m_k = E[q_{ik}]$ which represents subpopulation $k$'s relative size, $N_k/N$.
All of these models assume a Binomial likelihood, which most Bayesian probabilistic programming languages expect to be parameterized by an integer-valued $d_i$.  Additionally, the random degree and transmission bias models require an integer-valued parameter for each subpopulation size.  These integer-valued parameters make implementation in Stan difficult.
Current implementations of these models have varied in their inference algorithms and in their estimation of the integer-valued parameters with some using fully hierarchical specifications (with a noninformative prior and estimated via a Metropolis-Gibbs algorithm, as in \citet{maltiel2014NSUM}), and others using a constrained version of the model which treats the subpopulation sizes as known \citep{bojanowski2025stansum}. The barrier effects model has been implemented in Stan by evaluating the loglikelihood directly \citep{baum2023uses,bojanowski2025stansum}.}
\citet{teo2019estimating} introduced a related model, allowing the
incorporation of demographics about the ARD sample as regression coefficients.
\citet{laga2023correlated} introduce a further extension which allows 
for correlation in the bias between groups. Some recent work has focused on the 
performance of NSUM estimators under different simulated data conditions 
\citep{diaz2025performance,kunke2024comparing,lubbers2024measurement}.
\citet{laga2021thirty}
provides a detailed review of existing NSUM models, including a discussion
of Bayesian methods specific to NSUM. {\color{black}In this work we 
aim to provide general guidelines for any Bayesian model considered for ARD, including those that may not capture structure within the NSUM framework.}

{\color{black}\section{Simulating ARD}\label{sec:sim_data}}

Having reviewed several existing Bayesian models for ARD we can 
now describe
the data we will use to fit and compare these models. 
Here we will consider synthetic ARD, 
containing realistic complex structure.
{\color{black}This allows these examples to be completely self contained
and provides an opportunity to work with ARD models without having to
deal with the complexities of real data initially.}
We will first simulate data from the model proposed in 
\citet{mccormick_latent_2015}.
We will then create 
a dataset based on the real parameters of 
{\color{black}the classical dataset introduced by}
\citet{mccarty2001comparing}.
This will allow us to compare a range of Bayesian
models in the setting where we know the true parameter values.
{\color{black}This provides a clear framework to examine the
performance of increasingly complex ARD models to this data,
both in terms of metrics which can be used for real data, while also
comparing to true characteristics which are only known in simulation.}
Using simulated data also allows
readers to replicate the examples in this paper exactly and creates
realistic data from which readers can design their own experiments.

{\color{black}Given these simulated datasets, we will then 
describe the steps required to fit the models of
Section~\ref{mod-null}-\ref{model-latent} to each of these datasets,
outlining the key steps in fitting these models.}
All code to reproduce these ARD datasets is available
in our associated Github repository 
\url{https://github.com/OwenWard/ARD_Review}.
\\
\subsection{Synthetic data from the latent space model}
We first simulate data where the true network structure is that
considered in \citet{mccormick_latent_2015}.
Each edge in the underlying network, $A_{ij}$,
is generated
via a log-linear model
$$
\mathbb{E}(A_{ij}|g_i, g_j, \zeta, \mathbf{z_i}, \mathbf{z_j}) =
\exp(g_i + g_j + \zeta \mathbf{z_i}^T \mathbf{z_j}).
$$
Here the latent positions are drawn uniformly on the $p+1$ dimensional hypersphere.
We consider $K=15$ subpopulations and randomly place the centers of
these subpopulations in the latent space, before randomly assigning some nodes
from the population
to
each subpopulation of 15 subpopulations,
as a function of their distance to the subpopulation.
To match realistic applications
of ARD models, we will assume the true size of
the first 14 subpopulations are known, with interest
in inferring the remaining unknown subpopulation, along with the individual
node degrees.

We then take a sample of nodes
of size 1000
(1\% of the 
population) and simulate the number of edges they share
with nodes in the specified subpopulations,
given the true latent parameters. We show the 
true degree of the nodes in this sample in Figure~\ref{fig:latent_true_degree}.
With this simulated data we will fit each of our null models, 
along with the models of \citet{zheng_how_2006}
and the model of \citet{mccormick_latent_2015},
which corresponds to fitting 
the true model.

\subsection{A synthetic replicate of  \citet{mccarty2001comparing}}
\begin{figure}[ht!]
	\centering
    \begin{subfigure}{0.49\textwidth}
        \centering
        \includegraphics[width=\textwidth]{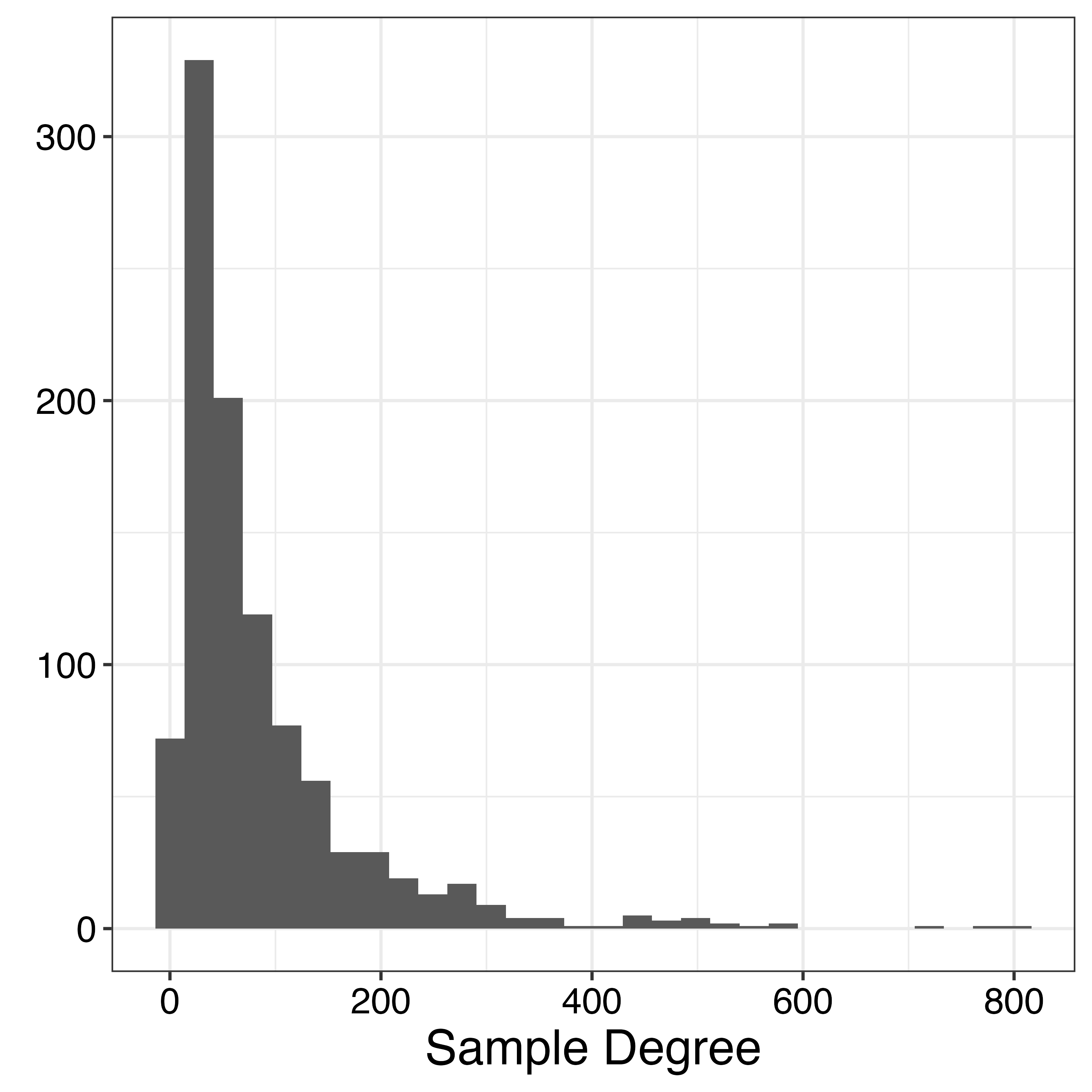}
        \caption{Synthetic Latent Space model data.\\}
        \label{fig:latent_true_degree}
    \end{subfigure}
    \begin{subfigure}{0.49\textwidth}
        \centering
        \includegraphics[width=\textwidth]{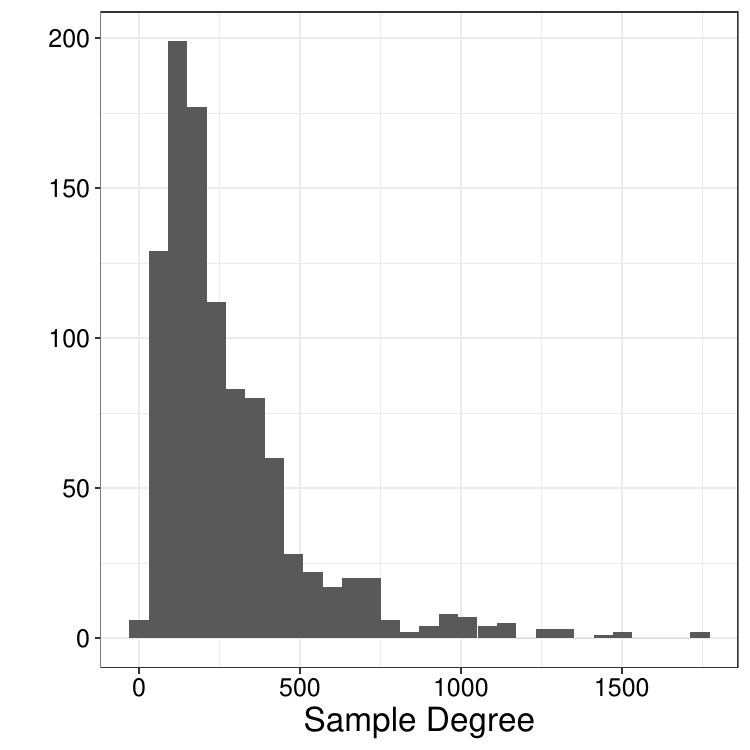}
        \caption{Replicate \citet{mccarty2001comparing} data.}
        \label{fig:mix_true_degree}
    \end{subfigure}
	\caption{Distribution of the true degree of the nodes in ARD sample for 
    each of the synthetic datasets used.}
	\label{fig:sim_true_degree}
\end{figure}

The survey data collected by \citet{mccarty2001comparing} has become a 
benchmark data set for ARD analyses, 
{\color{black}and was made publicly available by \citet{feehan2016generalizing} along with}
parameter estimates for an ARD model developed by 
\citet{maltiel2015estimating} available in the \texttt{NSUM} package for 
\texttt{R} \citep{maltiel2014NSUM}. We simulate synthetic ARD under 
\citet{maltiel2015estimating}'s barrier effects model, with these parameter 
values, which replicates the data structure and many important features of 
real world ARD. We show the 
true degree of the nodes in this sample in Figure~\ref{fig:mix_true_degree},
{\color{black}which gives us an additional ground truth we can
compare the estimates from each of our models against}.
With this replicated data we will fit each of our null models, 
along with the models of \citet{zheng_how_2006}
and the barrier effects model from \citet{maltiel2015estimating}, which corresponds to fitting 
the true model.

\section{Fitting Bayesian ARD Models} \label{sec:fitting}
{\color{black}A key consideration when selecting a Bayesian ARD model for a specific application is the choice of inference method used to estimate the posterior distribution. Different approaches offer varying trade-offs between computational efficiency and the accuracy of the resulting posterior estimates. This choice is also related to the 
expressivity of the 
chosen model. 
In Section~\ref{sec-fitting}, we list the primary inference methods used to fit Bayesian ARD models and describe in detail the most popular modern approach, Hamiltonian Monte Carlo via the probabilistic programming language Stan.
We review appropriate syntax and formatting for representing a Bayesian ARD model in Stan, using the null model as an example, previewing the coherent collection of Stan code implementations for each of the models described in Sections 3.1-3.3.
Regardless of the inference method used, 
most Bayesian ARD models assume a complex latent variable structure which requires that priors are suitably chosen and identifiability issues are appropriately addressed. 
We discuss these challenges and several approaches from the literature in Section~\ref{sec-iden}. 
When performing Bayesian inference via any MCMC method, it is important to verify that the Markov Chain has converged before the posterior samples are used for inference. We briefly describe how general MCMC diagnostics translate to the context of ARD models and identifiability in Section~\ref{sec-conv}.}


\subsection{Fitting algorithms}
\label{sec-fitting}
A range of computational methods have been considered to fit Bayesian ARD 
models.
Markov Chain Monte Carlo (MCMC) has been widely used for posterior estimation,
often with specialized code for an individual model. \citet{zheng_how_2006}
constructed a Gibbs-Metropolis algorithm, tuning the Metropolis jumping
scales to obtain the desired acceptance probabilities.
\citet{mccormick_how_2010} extended this approach to the nonrandom 
mixing model, with a similar inference procedure used in
\citet{mccormick_latent_2015}.

\citet{sahai2019estimating} utilise the Stan inference
platform \citep{stan2024}
to perform Bayesian inference. This uses the
no-U-turn sampler of \citet{hoffman2014no}
to perform an adaptive variant of Hamiltonian Monte Carlo.
Stan also provides a convenient programming language for specifying 
each component of the Bayesian model to be fit.
{\color{black}Stan has been used in many recent ARD models and provides 
practitioners with a convenient framework for users to fit these models, without
the need to construct their own Bayesian inference procedures. To make it easier for
users to fit these models,}
we provide a coherent collection of Stan code implementations 
for each of the models described in Section~\ref{mod-null}-\ref{model-latent}
that
we fit to data below.
{\color{black}
We note that \citet{stansum} provides an implementation of several recent ARD models also, which
overlaps with some of the models we have included here and complements this
discussion.}

To illustrate how Stan is used to fit Bayesian models, we include the Stan code to fit the null model with equal degrees in Code Block~1.
This is the simplest Bayesian ARD model, and we choose diffuse priors for the degree and group proportions for simple exposition. We discuss the role of choosing these priors and other important practical considerations, notably  rescaling estimates to match known subpopulations,
in more detail {\color{black}in Section~\ref{sec-iden}}.
Stan model code is separated into blocks for the data, parameters, and model. Other blocks may be 
defined; for example, in the following section, we will add a transformed parameters block to incorporate 
rescaling and in Section \ref{sec:checking}, we will add a generated quantities block to draw replicated 
data from the posterior predictive distribution. The data block defines a matrix of ARD observations, 
$Y$, with $N$ rows (for egos) and $K$ columns (for subpopulations).  The parameter block defines the 
unknown latent variables or parameters to be estimated for this model: $log(d)$, the common degree or 
personal network size (on the $log$ scale), and the $\beta_k$'s, which are subpopulation prevalence 
parameters, for $k=1, \dots K$.  The model block includes distributional assumptions for the data (i.e., 
the Poisson likelihood) and model parameters (i.e., the diffuse Normal priors).  Our public GitHub 
repository contains Stan files for all the primary Bayesian ARD models discussed in 
Section~\ref{mod-null}-\ref{model-latent}.

\begin{lstlisting}[caption={Stan code for Null Model with Common Degree}]
// Stan Model for the Erdos Renyi Model
data {
  int<lower=0> N;
  int<lower=0> K;
  array[N, K] int y;
}
parameters {
  real log_d; // the common population degree
  vector[K] beta;
}
model {  
  log_d ~ normal(0, 25);
  beta ~ normal(0, 5); 
  real exp_log_d = exp(log_d);
  for (n in 1:N) {
    y[n] ~ poisson(exp_log_d * exp(beta));
  }
}

\end{lstlisting}

Other advanced 
inference procedures have also been considered to perform
Bayesian inference when modeling ARD. \citet{jones2021scalable}
propose a variational inference scheme to model ARD where 
the underlying network is generated from a
mixed membership stochastic block model \citep{airoldi2008mixed}.
Variational inference (VI) is an alternative approach to doing Bayesian
inference with substantial differences to 
MCMC approaches. Instead of collecting samples from
the true posterior distribution using MCMC, 
VI seeks to find a class of distributions which can approximate
the true posterior. Thus VI is well suited to problems where 
massive datasets are available and more traditional methods 
may not be computationally feasible \citep{blei2017variational}, 
{\color{black}but it may generate samples that do not accurately reflect the 
true posterior and underlying uncertainty.}

\subsection{Prior distributions and scaling} 
\label{sec-iden}
{\color{black}The first} challenge a practitioner may encounter when attempting to fit a Bayesian model is how to choose appropriate priors.
In practice, many of the prior distributions for parameters in these models can be  specified as vague or noninformative distributions, using conjugate relationships when possible.
However, the priors used can also incorporate existing 
information about the parameters of interest, if available.
In \citet{zheng_how_2006}, normal prior distributions are
used for both $\alpha_i$ and $\beta_k$, with uniformative priors
used for the parameters in these priors.
The model of \citet{mccormick_latent_2015}  assumes that individuals’ 
latent positions
are uniformly distributed across the sphere, which {\color{black}naturally} reflects the 
random sampling assumption
for survey respondents (i.e., that any population member has equal 
probability of joining our sample
data). Latent positions for the members of each subpopulation are 
distributed according to the
von Mises-Fisher distribution, which is related to the multivariate Normal 
distribution and is a
common choice for points on a sphere \citep{mardia_2009}. To 
address identifiability issues in latent space models,
the centers of a subset of subpopulations are fixed in advance.
The authors recommend using diverse subpopulations with known 
characteristics. For example, in \citet{mccormick_latent_2015},
the estimated latent center of the
subpopulation of individuals with AIDS, which is most common among young 
males, is close to
the fixed center of the subpopulation with the name Christopher, which SSA 
data tells us is also
common among young males.
In cases where prior specification is particularly important,
diagnostic tools, such as prior predictive checks, can be used to investigate and evaluate their suitability \citep{gabry2019visualization}.

{\color{black}
While we considered identifiability for the Latent Space model above, none of the ARD models in Section~\ref{sec:models}} are fully identifiable {\color{black}without additional constraints.}
We can only reliably recover personal network sizes and subpopulation sizes up to a multiplicative scaling factor.  
To see this more clearly, consider the simple model given in Code Block~1.
If the degree parameter was multiplied by some positive constant, dividing the subpopulation proportions by the same constant would result in the same overall rates (and the same likelihood).  
Fortunately, incorporating information on the sizes of known subpopulations allows us to rescale the model parameters back to the appropriate scale, providing interpretable estimates of both individual degrees and subpopulation sizes.
Accounting for this has been approached in multiple ways 
in the Bayesian ARD literature. \citet{zheng_how_2006}
do this by fitting their model with no constraints,
rescaling their estimates afterwards such that the subpopulation
estimates for rare populations agree with their true population
prevalence (which is known in the case of rare names).
\citet{laga2023correlated} provide an alternative general approach 
to this problem, given an NSUM model which incorporates correlation
between subpopulations. We consider a modified version 
of this approach, for models without
this correlation.
To incorporate this scaling in our null model
requires minimal modifications of the Stan code, by
providing the total prevalence of several known subpopulations
in the population.
Adding this additional data requires a minor modification
of the model fitted through
adding a `transformed parameters' block before specifying the fitted
model, as
is shown in Code Block~2.

\begin{lstlisting}[caption={Stan code for Scaling Null Model with Common Degree}]
// Stan Model for the Erdos Renyi Model with scaling
data {
  int<lower=0> N;
  int<lower=0> K;
  array[N, K] int y;
  int<lower=0, upper=K> n_known;

  // total prevalence for a subset of known subpopulations
  array[n_known] int<lower=1,
  upper=K> idx;
  real<lower=0, upper=1> known_prev; 
}
parameters {
  real log_d; // the common population degree
  vector[K] beta;
}
transformed parameters { // rescaling
  vector[K] scaled_beta;
  real scaled_log_d;
  real C;
  C = log(sum(exp(beta[idx])/known_prev) );  
  scaled_log_d = log_d + C;
  scaled_beta = beta - C;
  vector[K] b = exp(scaled_beta);
}
model { 
  log_d ~ normal(0, 25);
  beta ~ normal(0, 5); 
  real exp_log_d = exp(scaled_log_d); // use the rescaled degrees
  for (n in 1:N) {
    y[n] ~ poisson(exp_log_d .* b);
  }
}
% \end{lstlisting}

This rescaling procedure constrains the total size of several known
subpopulations to match their true total size, providing interpretable 
estimates for the node degrees and the other unknown subpopulations
of interest. This allows examining these estimates directly, 
and also aids in Bayesian model convergence, as discussed below.
In the appendix we provide the exact priors for each of the four models we fit 
in Stan.
{\color{black}We have incorporated this scaling within our associated Stan models to provide
users with an easy to use procedure, given the likely available information.
In practice, appropriately scaling these estimates can {\color{red} be performed as a post-hoc process, but the procedure can} be problem specific and may require the use of additional information. 

\subsection{Bayesian convergence checks}
\label{sec-conv}
A challenge which is common to all Bayesian modeling 
is how to assess when an algorithm to 
estimate a posterior distribution has converged.
Several metrics have been developed for
MCMC methods, the mostly widely used being 
the $\hat{R}$ metric originally
proposed by \citet{gelman1992inference}. 
Several (commonly 4) Markov Chain Monte Carlo procedures 
are run in parallel from different initial values and convergence is assessed in terms
of the within chain and between chain variability
of the estimates for each parameter in the model.
\citet{zheng_how_2006, mccormick_how_2010,mccormick_latent_2015} each use this metric, 
considering acceptable convergence to be when $\hat{R}<1.1$
for all parameters. We note that the current documentation
of Stan \citep{stan2024} recommends only using posterior samples when
$\hat{R}<1.05$.

Stan automatically computes this convergence metric
for each parameter in the model, along with measures of the effective
sample size \citep{gelman2013bayesian}. This convergence 
can also be assessed graphically using traceplots \citep{bayesplot},
and 
we will consider more advanced diagnostics in the following section.
We note that if we fit the null models or the models of
\citet{zheng_how_2006,mccormick_latent_2015} without this scaling 
step, {\color{black}leaving all model parameters unconstrained,}
we obtain large values for $\hat{R}$ for all parameters, 
indicating our Markov chains struggle to converge. Incorporating
scaling leads to good convergence metrics for all models.
We show this in Figure~\ref{fig:sim_trace}, fitting the Erdos Renyi 
model to the Latent Space simulated data. Without scaling, the trace plots
indicate issues with convergence of the MCMC algorithm, which is also 
seen in the $\hat{R}$ diagnostic. {\color{red}Convergence can also be achieved if certain parameters, such as the
mean of the prior for the log degree, are forced to be 0 \citep{zheng_how_2006}.}

\begin{figure}[ht!]
	\centering
    \begin{subfigure}{0.49\textwidth}
        \centering
        \includegraphics[width=\textwidth]{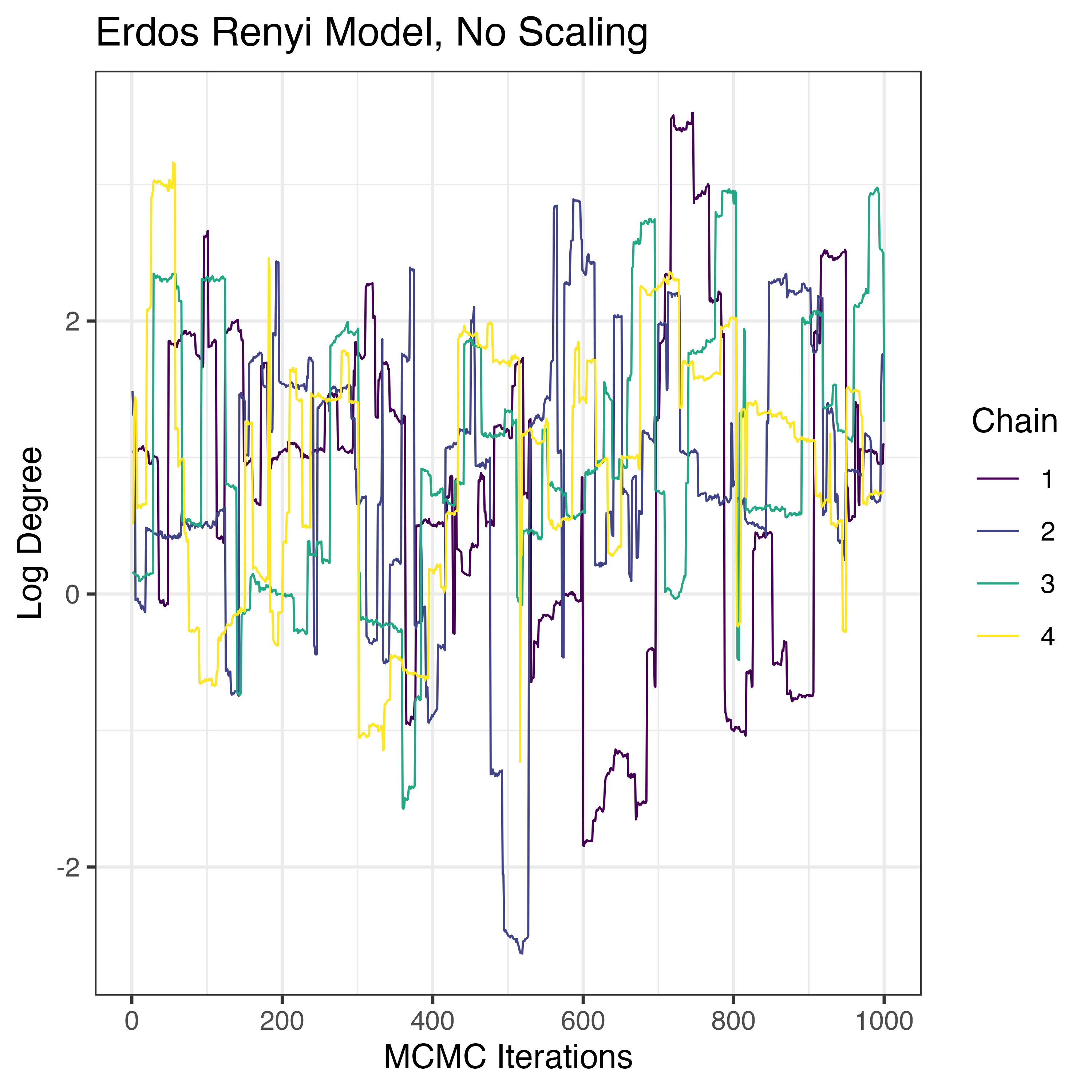}
        \caption{Trace plot without scaling, $\hat{R}=1.10$.}
        \label{fig:latent_trace_no_scaling}
    \end{subfigure}
    \begin{subfigure}{0.49\textwidth}
        \centering
        \includegraphics[width=\textwidth]{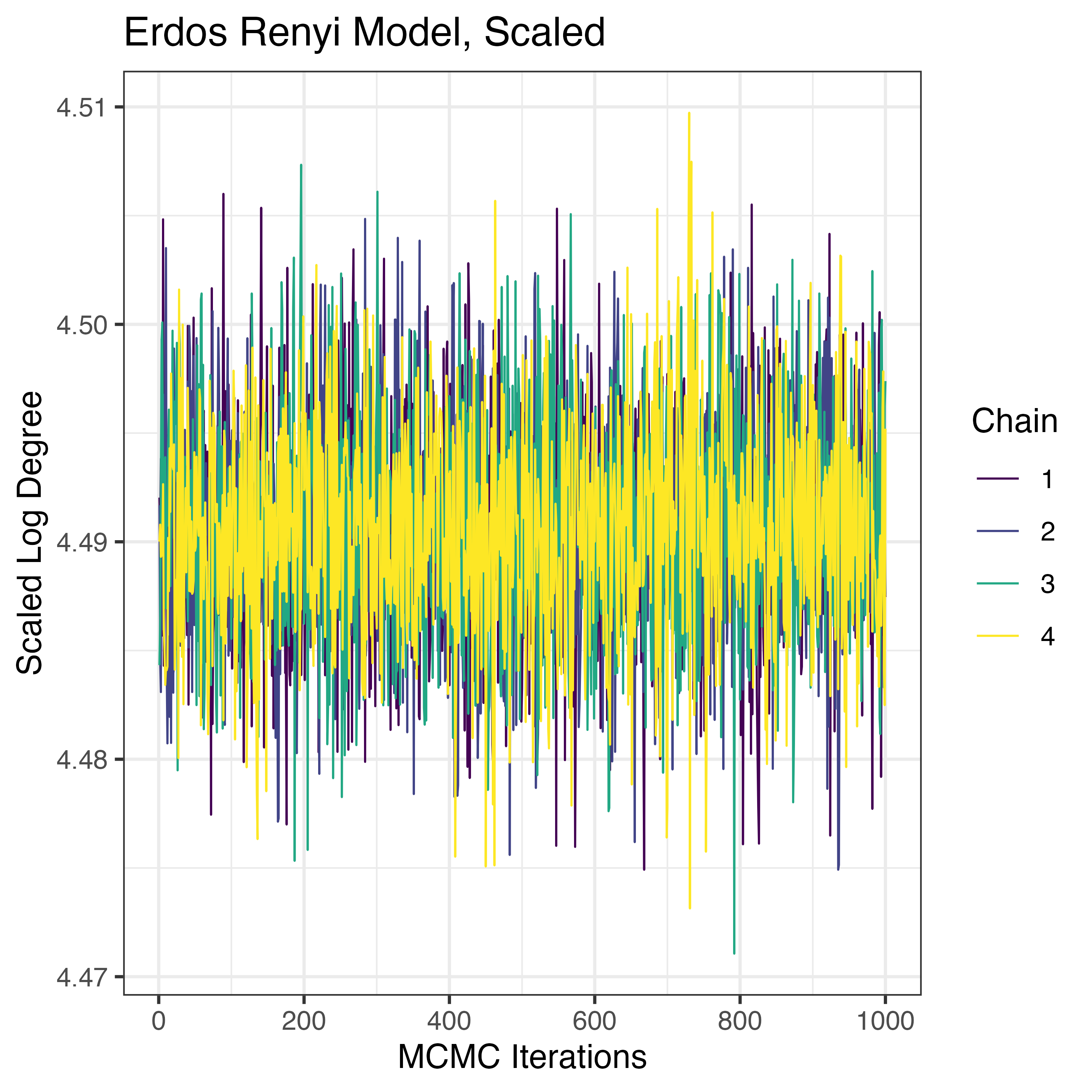}
        \caption{Trace plot with scaling, $\hat{R}=1.00$.}
        \label{fig:latent_trace_scaling}
    \end{subfigure}
	\caption{Trace plots of the posterior draws of the log-degree
    parameter in the Erdos Renyi model, with and without scaling.
    The trace plot without scaling shows poor convergence, indicating
    the model fails to fit well. This is seen 
    in the associated values of $\hat{R}$.}
	\label{fig:sim_trace}
\end{figure}

\section{Model Checking and Comparison}
\label{sec:checking}
An important component in the use of any Bayesian
statistical model is the ability to assess how appropriate the
fitted model is, given the observed data.
Bayesian model checking is an active area of
research 
and modern Bayesian models are often
constructed by iterating through this
process to determine a final model \citep{gelman2020bayesian, gelman2013bayesian}.
While several Bayesian models have been proposed for ARD,
model checking for these methods is limited. 
In Section~\ref{sec-recov} we take advantage of 
our use of simulated data to examine how several of the models
described in Section~\ref{sec:models} recover the true degree distributions
and unknown subpopulation sizes. While this is not possible 
for real data, it serves to highlight the tradeoffs between models
of different flexibility. In Section~\ref{sec-checking} we review the
limited existing approaches in the literature for ARD model checking.
Finally, in Section~\ref{sec:model-sel} we consider model
selection in the context of Bayesian ARD models. We discuss the
challenges presented in this context and
{\color{red}highlight some previously undocumented issues with regards to 
the application of cross-validation to Bayesian ARD models.}


\subsection{Recovering degree and subpopulation estimates}
\label{sec-recov}
With the two synthetic datasets described in Section~\ref{sec:sim_data}
we 
fit each of the two null models, and the model of \citet{zheng_how_2006}.
We then fit the appropriate true model for each
of these datasets.
We can then assess the performance of each of these 
models in terms of recovering the true degree distribution
of the nodes in our ARD, along with the estimated sizes
of the unknown populations. While these are both important goals
with real data, the true values of these quantities would be unknown in
practice.  However, generating data from ARD models and assessing model fit can provide a better understanding of the statistical properties of ARD and its models.  All the code used to evaluate the fit of these models to our synthetic datasets is available in our public GitHub repository; tweaking the parameter values or data sizes to better represent a given ARD data setting allows for the assessment of model performance and parameter recovery in other specific settings {\color{black}of interest to users.}

We first examine the estimated degree distributions
under each of the four models we fit to each dataset, with histograms of the true simulated degrees and smoothed densities of each model's estimated degree distribution shown in Figure~\ref{fig:sim_all_true_degree}.  The Erd\"{o}s-R\'{e}nyi model's degree estimates are represented as vertical lines (point masses) in this plot as the model assumes a common degree across all participants. 
For both simulated datasets, this model tends to overestimate the true degrees.
For the Latent Space simulation, the degree estimates from fitting the overdispersed model more closely match the true degree distribution than those obtained when the true model is fitted. 
{\color{black}This highlights the inherent challenges of fitting such a complex hierarchical
model, which also overestimates the size of the unknown subpopulation 
(which is common to all models fit).
We note that this could also
be related to the choice of scaling procedure used, highlighting how different
scaling methods may be needed in different settings. 
\citet{mccormick_latent_2015} also 
fixed the centers of a subset of their target populations to aid with
identifiability. In practice, fitting such a complex model
may prove excessively challenging.}
For the synthetic \citet{mccarty2001comparing}
data the varying degree model provides the best fit to the true degrees, while all other models, including the true barrier effects model, tend to overestimate degree.

\begin{figure}[ht!]
	\centering
    \begin{subfigure}{0.49\textwidth}
        \centering
        \includegraphics[width=\textwidth]{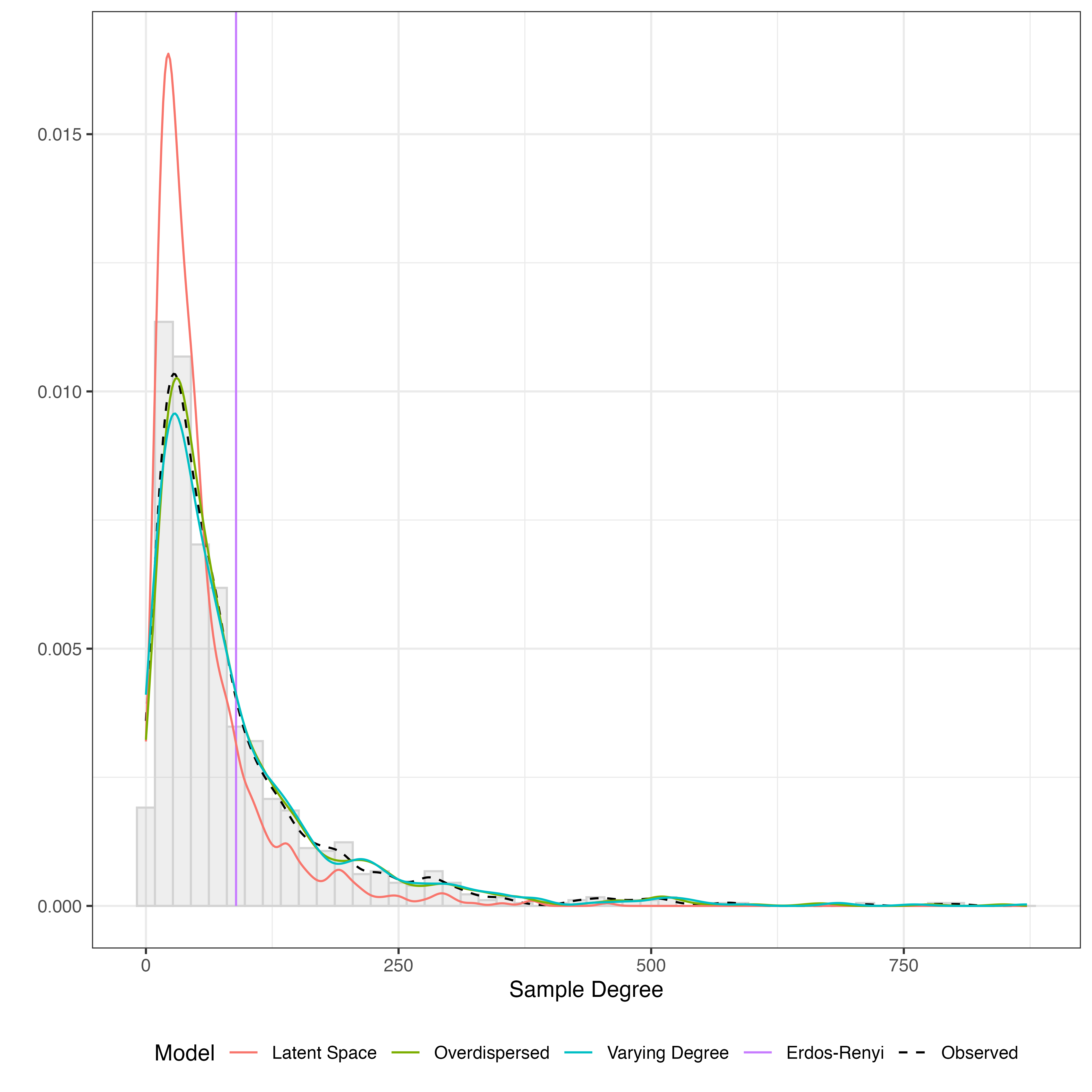}
        \caption{Synthetic Latent Space model data.\\}
        \label{fig:latent_all_degree_2006_2015}
    \end{subfigure}
    \begin{subfigure}{0.49\textwidth}
        \centering
        \includegraphics[width=\textwidth]{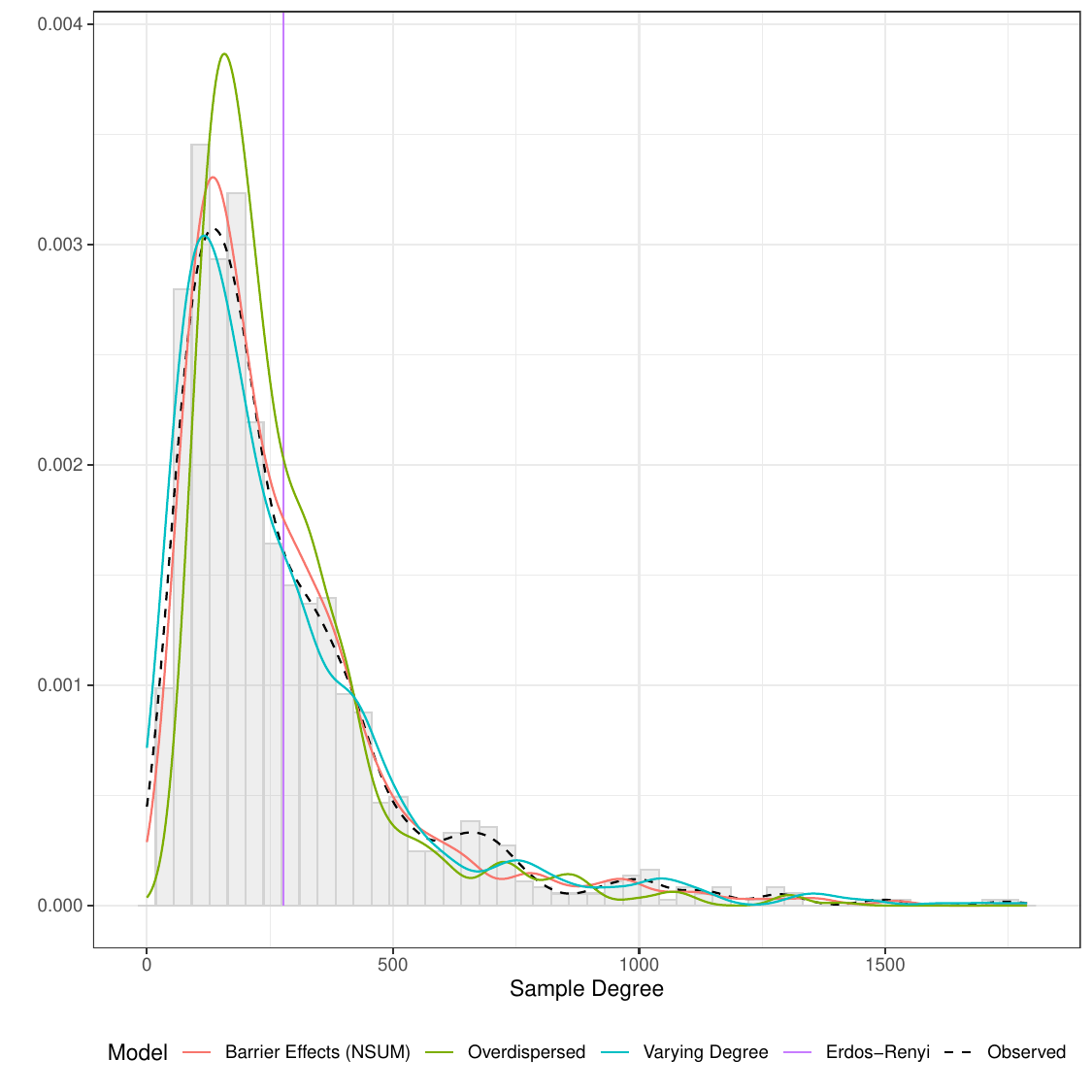}
        \caption{Replicate \citet{mccarty2001comparing} data.}
        \label{fig:mix_all_degree}
    \end{subfigure}
	\caption{True degree distribution and posterior degree distribution under common ARD models for 
    each of the synthetic datasets examined.}
	\label{fig:sim_all_true_degree}
\end{figure}

Similarly, we can assess these models in terms 
of their ability to estimate the size of 
both known and unknown subpopulations.
We first fit the overdispersed model for the synthetic \citet{mccarty2001comparing} data,
and show the estimated size of all 30 subpopulations, after scaling, 
in Figure~\ref{fig:McCarty_est_subpop}. We see that the posterior intervals
for most subpopulations agree with the true size.

\begin{figure}[ht!]
	\centering
   \includegraphics[width=\textwidth, height=\textwidth]{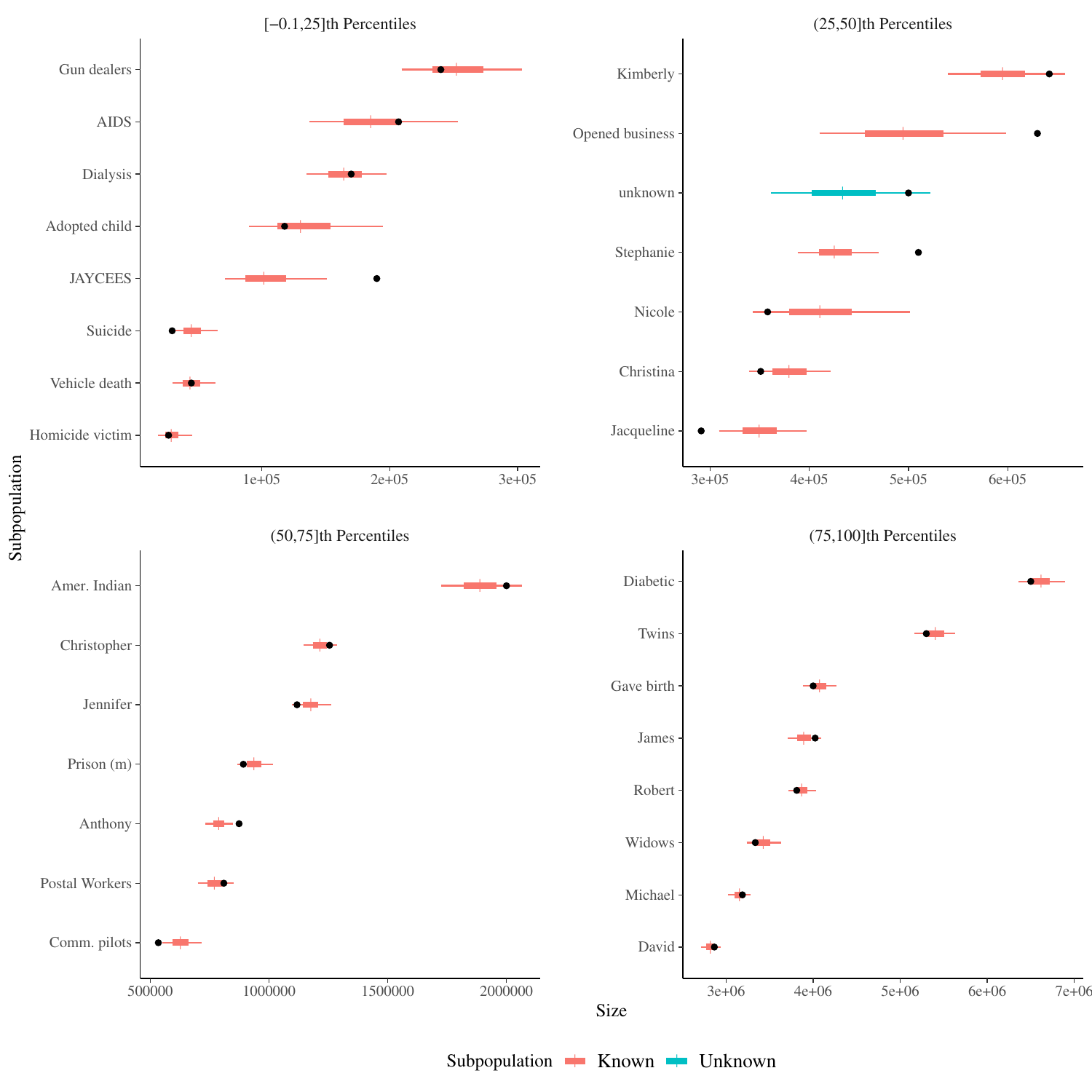}
	\caption{Recovery of subpopulation sizes for all 
    subpopulations under the \citet{zheng_how_2006} model for the replicate \citet{mccarty2001comparing} data.  Black points indicate the true size of each subpopulation.  Colored points (red for known subpopulations and blue for the unknown subpopulation) indicate posterior medians, with thin and thick colored bars representing 90\% and 50\% credible intervals respectively.  Subpopulations are ordered by their posterior median size.}
	\label{fig:McCarty_est_subpop}
\end{figure}

As ARD models are commonly used for inferring the size of 
one (or more) unknown subpopulations, we wish to compare each
of the models we fit to our synthetic datasets in their ability
to recover unknown subpopulations. In Figure~\ref{fig:sim_subpop_all}
we show the estimated size of the unknown subpopulation in each dataset according to each model.
For the latent space data, 
{\color{black} the true model shows similar (poor) 
performance to the other models considered, with all showing 
clear overestimation. This again highlights the challenge 
of performing an appropriate scaling for these models, particularly
when there are many latent variables.}
For the synthetic \cite{mccarty2001comparing} data the true model 
shows the best performance, but all models underestimate the unknown subpopulation size. Underestimation is a bit more severe for the simpler models, but the overdispersed model's uncertainty better matches the uncertainty under the true model.

In some settings we may care more about estimating one of these parameters: personal network sizes or (unknown) subpopulation sizes. Based on the synthetic datasets we examine here, the varying degree model's estimated degree distribution closely matches the true degrees across both datasets, although the overdispersed model provides better fit for the Latent Space model data. For the size of the unknown subpopulation, the true model 
can provide the best estimates, {\color{black}but all models have some issues.}
In the next section, we discuss how to assess model fit in the more practical setting where the true parameters are unknown.

\begin{figure}[ht!]
    \centering
    \begin{subfigure}[t]{0.48\textwidth}
        \centering
        \includegraphics[width=\textwidth]{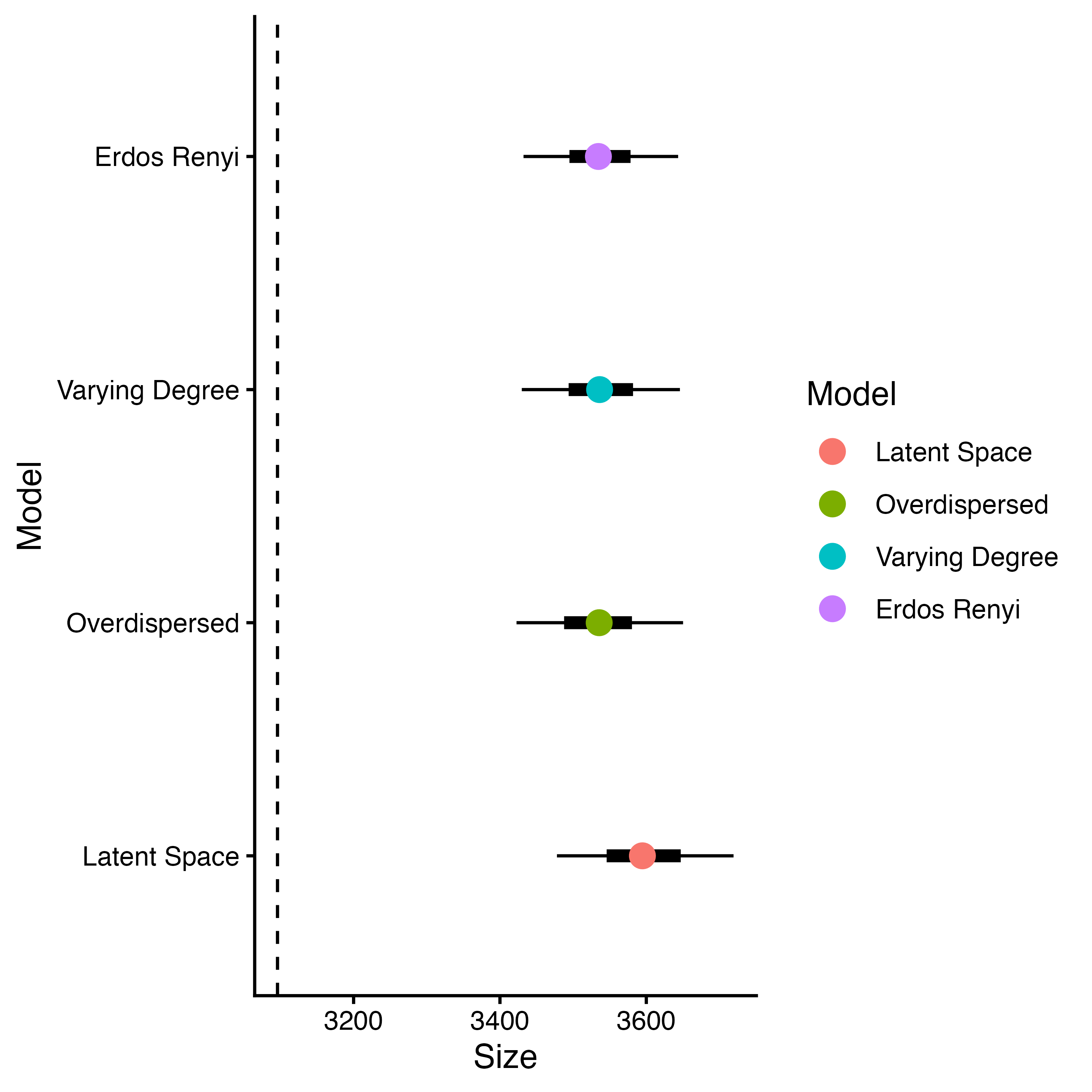}
        \caption{Unknown subpopulation estimation for synthetic Latent Space model data. 
        }
        \label{fig:latent_subpop}
    \end{subfigure}
    \hfill
    \begin{subfigure}[t]{0.48\textwidth}
        \centering
        \includegraphics[width=\textwidth]{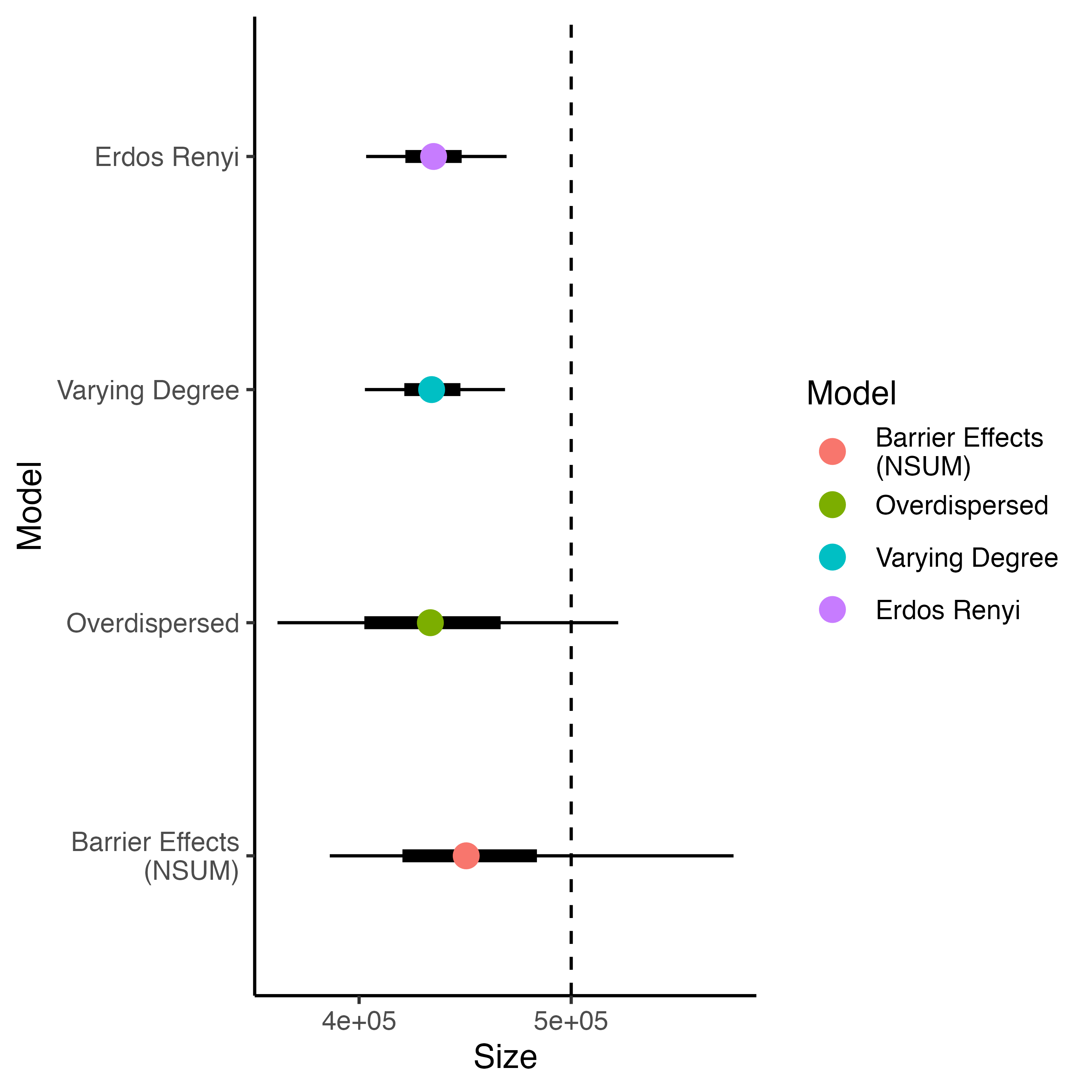}
        \caption{Unknown subpopulation estimation for replicate \citet{mccarty2001comparing} data. 
        The 90\% posterior intervals for the true model and the overdispersed model both contain the true value.}
        \label{fig:mix_subpop}
    \end{subfigure}
    \caption{Subpopulation estimates for the unknown subpopulation under common ARD models for 
    each of the synthetic datasets examined. For each model
    we show posterior means along with 50\% and 90\% posterior intervals.}
    \label{fig:sim_subpop_all}
\end{figure}

{\color{black}\subsection{Model Checking for ARD}
\label{sec-checking}}
Up until recent work \citep{laga2023correlated, park2021segregated},
the only existing procedure for model checking in Bayesian 
ARD models was the utilisation of posterior predictive checks 
by \citet{zheng_how_2006}.
Posterior predictive checks (PPCs) \citep{gelman2013bayesian}
are a general tool which can be applied to any Bayesian model 
where data can be simulated from the fitted model, using the posterior
draws as parameter values for these simulations. \citet{zheng_how_2006}
incorporated these by obtaining a replicated sample draw
of $\tilde{y}_{ik}$ for each draw from the posterior distribution for each parameter.
For each posterior predictive draw, a $N \times K$ matrix, $\tilde{y}$, they then compute the proportion of entries in this matrix, $\tilde{y}_{ik}$, equal to potential observed values ($m = 1, 2, 3, \dots$ and compare it to the true proportion in
the original data.  This corresponds to checking how well replicated data simulated under your resulting posterior distribution matches the original observed data, in terms of the frequency of values in the ARD matrix (i.e., how many respondents report knowing exactly $1,2,3\dots$ members of each subpopulation).

Using Stan to perform Bayesian inference allows us to obtain posterior 
predictive draws at the same time as fitting the original model, through the 
use of a ``generated quantities" block in the Stan file, after the ``model" 
block. 
In Code Block~3 we show the corresponding generated quantities block for the 
Erd\"{o}s-R\'{e}nyi model shown previously. This block creates a new matrix 
\texttt{y\_sim} which contains a single draw from the posterior predictive 
distribution.  Just like the parameters defined in earlier blocks of our Stan 
file, the output from our MCMC sampling algorithm will contain a draw of $
\tilde{y}$ for each draw from the posterior (e.g., if we run our MCMC sampler 
for $I$ iterations (after a burn-in or warmup period), our output will contain 
a vector with $I$ entries (draws) of $\beta$ and an array with $I$ draws of 
the $N \times K$ matrix of $\tilde{y}$). To simulate draws from the model's 
likelihood, given parameter values from our posterior distribution (which Stan 
plugs in automatically), we use the \texttt{\_rng}, random number generator, 
version of the Poisson distribution, here on the log scale.

\begin{lstlisting}[caption={Stan generated quantities block for Null Model with Common Degree}]
// previous code same as shown above...
model {  
  log_d ~ normal(0, 25);
  beta ~ normal(0, 5); 
  real exp_log_d = exp(scaled_log_d);
  for (n in 1:N) {
    y[n] ~ poisson(exp_log_d * exp(beta));
  }
}

generated quantities {
  array[N, K] int y_sim;
  real curr_log_d = scaled_log_d;
  for (n in 1:N) {
    for(k in 1:K){
      y_sim[n, k] = poisson_log_rng(curr_log_d + scaled_beta[k]);
    }
  }

}
\end{lstlisting}

Given sample draws of $y_{ik}$ from this posterior predictive
distribution, \citet{zheng_how_2006} then computed the proportion 
of times $y_{ik}=m$ within a specific subpopulation, comparing this
estimate to the true proportion. If these simulated 
proportions agree with the true proportion, it is an indication that 
the fitted model is successfully capturing aspects of the true data.

We illustrate the results of these PPCs for our simulated
datasets and the models we have fit for $m=0,1,3,5,10$. In Figure~\ref{fig:latent_ppc_all}
we show these checks for each of the four models fit to the synthetic 
latent space data while in Figure~\ref{fig:mccarty_ppc_all} we repeat this
for the synthetic \citet{mccarty2001comparing} data. In each case we show the 95\% posterior
intervals
of the proportion of times $y_{ik}=m$ for this range of values of $m$, with
a separate interval for each subpopulation. If these intervals agree with 
the true proportion in the data then they should overlap the dashed gray line.
We show posterior intervals which do not contain the true value in red (and
those which do in blue) in each plot. As we increase the complexity of
our fitted model we see that more of these intervals agree with the true value.
In each case the intervals under fitting the true model align most closely with
the real proportions. While this provides an important graphical tool to assess
model fit, in both cases we see large agreement in the PPCs from the true
model and the (simpler) overdispersed model.
In \citet{zheng_how_2006}, these posterior predictive checks for $m=9$ are especially useful, as they call attention to a ``heaping'' effect.
This effect commonly occurs in real data and describes the phenomenon that respondents are more likely to answer with round numbers (e.g., 5 or 10).  Since both our synthetic datasets are simulated directly from ARD models, they will not exhibit this behavior.  For real data, computing PPCs for near-round numbers (e.g., 9 or 11) can be particularly useful for diagnosing poor model fit.

\begin{figure}[ht!]
	\centering
   \includegraphics[width=\textwidth]{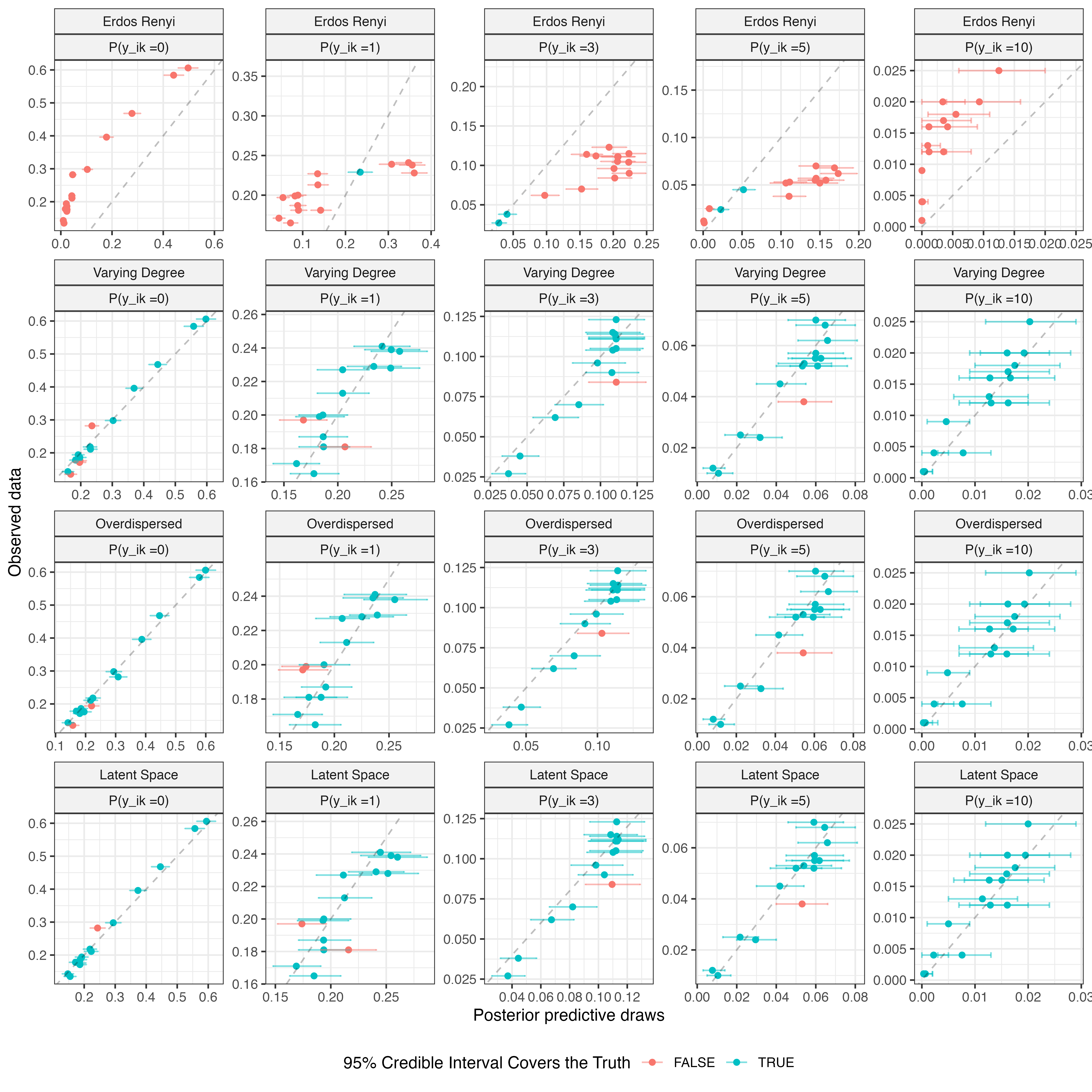}
	\caption{Posterior predictive checks for a range of values of $y_{ik}$ under common ARD models, fit to the synthetic Latent Space data.}
	\label{fig:latent_ppc_all}
\end{figure}

\begin{figure}[ht!]
	\centering
   \includegraphics[width=\textwidth]{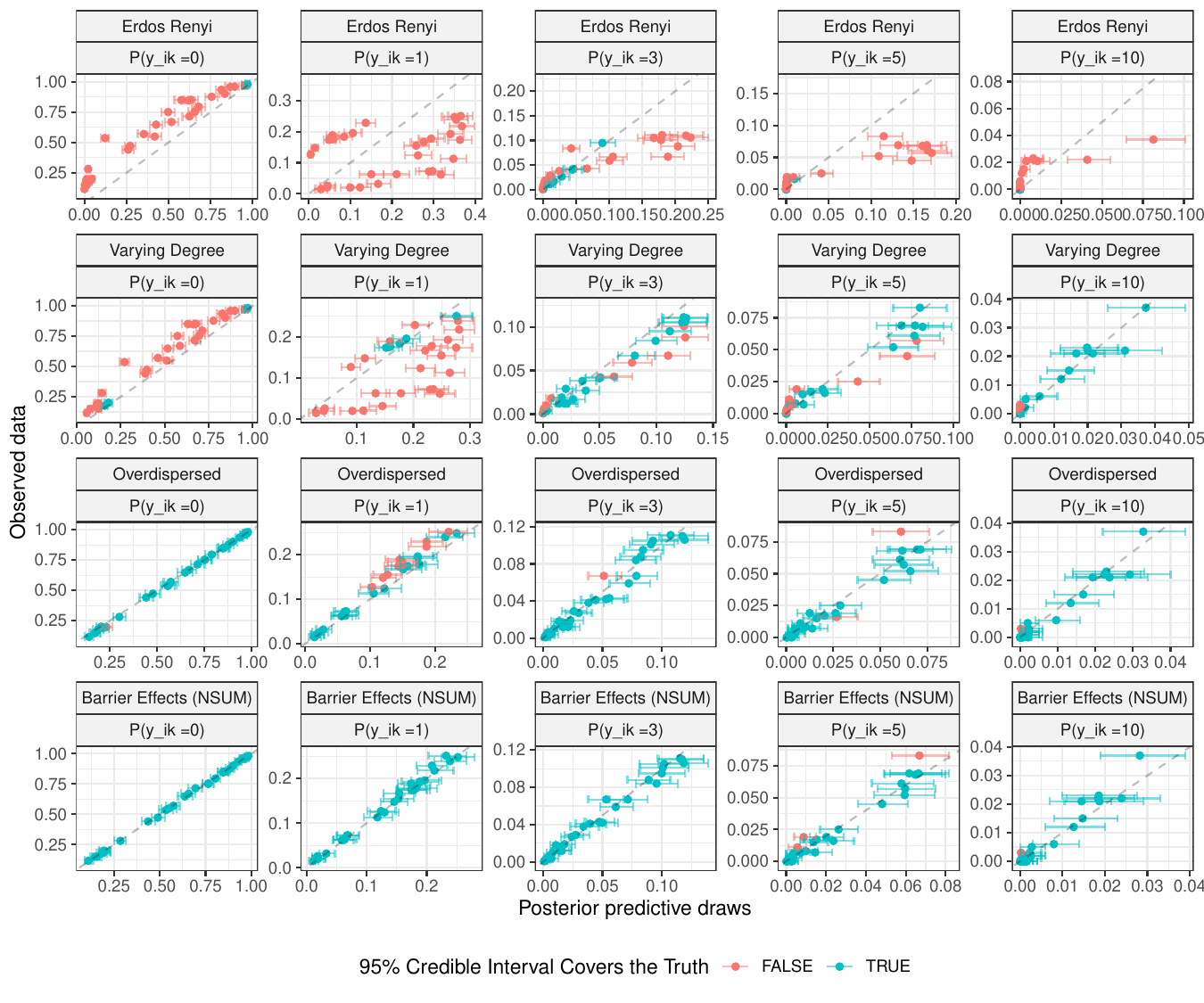}
	\caption{Posterior predictive checks for a range of values of $y_{ik}$ under common ARD models, fit to the replicate \citet{mccarty2001comparing} data.}
	\label{fig:mccarty_ppc_all}
\end{figure}

{\color{black}As highlighted above}, 
the recent work of \citet{laga2023correlated}
has also considered Bayesian model checking 
for fitting ARD models, proposing to use
surrogate residuals. These can be computed for Bayesian models
using samples from the posterior predictive distribution.
\citet{laga2023correlated} used these residuals to examine the 
impact of covariates in their Bayesian NSUM model. Like the PPCs 
considered above, these provide an important tool to assess if a fitted 
model 
is reasonable.

\subsection{Model Selection via Cross-Validation}
\label{sec:model-sel}

{\color{red} Cross-validation (CV) is a highly flexible approach for measuring predictive error and has been used to compare competing models in a range of data settings.
However, many of the details associated with implementing CV for ARD models have not been fully addressed in the current body of literature.
Given the structure of ARD data, it is important to note that multiple versions of cross-validation are possible: each fold can exclude (a) one entry, $y_{ik}$ (b) one individual or row, $y_{i1}, \dots y_{iK}$, (c) one ego group or subset of rows, in the case where demographic information is collected, or (d) one alter group, $y_{1k}, \dots y_{Nk}$. 
Each implementation allows for different kinds of model criticism and may be more or less useful.  
For example, excluding each individual entry, can be incredibly computationally expensive, especially for more complex models, as it requires fitting the model $N\times K$ times.

Implementing CV in general requires the ability to compute the log-likelihood for (some subset of) individual entries, $y_{ik}$, conditional on parameter values estimated from all other data. 
However, many of the ARD models discussed here include at least individual-level parameters and may also include entry-level parameters. 
This complicates the implementation of CV in the ARD setting.
For row-wise CV (version (b) above), when the $i$th row is heldout, the corresponding individual-level parameters (e.g., $d_i$) cannot be estimated from the training set and so the log-likelihood of the heldout $i$th row cannot be calculated, since the corresponding $d_i$ is missing. 
One could consider simply drawing $d_i$ from the prior, but this would make CV more sensitive to prior specification and less useful for model comparison.

Stan models come with a default implementation of leave-one-out (LOO) CV which uses Pareto smoothed importance sampling (PSIS) to approximate this quantity instead, which could potentially ease the computational burden \citep{vehtari2017practical}. These estimates have been used in recent work that considers ARD model comparison and selection quantitatively \citep{park2021segregated,baum2025explaining}.}
Stan's LOO-CV estimates are incredibly flexible but {\color{red}care must be taken in}
hierarchical models, which is particularly relevant here \citep{loo_package}.  
Hierarchical models generally include group-level parameters and removing all of the data for a given 
group will significantly impact the posterior distribution for that group-level parameter.  As a 
result, the PSIS approach usually doesn't work well in this setting, unless the group-level parameters 
can be integrated out \citep{vehtari2016bayesian, merkle2019bayesian}.  
{\color{red}
This has been documented in existing Stan case 
studies\footnote{See \url{https://users.aalto.fi/~ave/modelselection/roaches.html#cross-validation-checking}.}, including a Poisson regression model with a random intercept, where 
removing a single observation leads to large changes in the posterior for that observation.  This results in worryingly large estimates of $k$, the shape parameter of the Pareto distribution used to smooth the importance sampling weights, and warning messages from the \texttt{loo} package. The size of these estimates are useful diagnostics for the accuracy of the approximation; when $\hat{k}$ is too large, the asymptotics underlying the PSIS estimates can be slow to kick in and may not hold.  \citet{vehtari2016bayesian} warn that PSIS estimates are not reliable in these cases and that exact ($K$-fold) cross-validation or a more robust model should be pursued. 
In ARD models, the individual-level parameters (such as $d_i$) play the same role as the random intercept in the Poisson regression example. We found that small changes in the
prior distribution for the degree parameter could lead to dramatically different
LOO-CV estimates (each time resulting in large $\hat{k}$ estimates and producing warnings indicating they should not be used).
We encountered this issue 
when fitting a range of Bayesian ARD models discussed here and from the literature, in all but the simplest null model with a common degree parameter.

While PSIS approximations do not appear to ease the computational burden of CV, exact $K$-fold entry-wise CV (described as version (a) above) is a natural alternative.
As discussed above}, we need to be able to construct held out data subsets, for which we make predictions using a model fit on an additional nonoverlapping subset of the data.
As the models considered here contain node-level effects (the personal degree parameters, $d_i$),
which vary for each row of our ARD $Y$, we cannot use entire 
rows in the held out data fold.
Instead, we consider simply holding out random entries 
of $Y$. We partition the entries of $Y$ into $K$ folds, $Y^{1},\ldots,Y^{K}$.
Then for each fold $k=1,\ldots, K$ consisting of $Y^{k}$ 
we fit our model on all the entries of $Y$
except those in $Y^{k}$, before evaluating the held out model for the entries
in $Y^{k}$. 
This can be incorporated into the existing modeling framework using Stan,
which uses the expected log predictive density as the evaluation metric.
This approach can only be considered in the case of models which do not have 
{\color{red}entry-level} 
parameters for each entry of the ARD, $Y$. In particular,
this cannot be used to evaluate a model such as \citet{maltiel2015estimating}
or \citet{laga2023correlated}, {\color{red}as they would experience the same
issues which affect the PSIS approximation.}
Further work is needed to develop a more general
model selection procedure which can incorporate the vast flexibility 
present in modern Bayesian ARD models.

To demonstrate how a model selection procedure could be used for Bayesian 
ARD models, we consider applying this cross-validation approach
to the four models fit to the synthetic Latent space data as
an illustrative example. 
Using $K=10$ folds we fit each of our models and evaluate the predictive 
performance. To compare model fit, we estimate the expected log predictive density ($ELPD$) which is defined as the log density of all ARD entries, as an expectation over the population of all ARD samples, excluding the one we observe.  Since we only have access to a single sample of ARD, we estimate the $ELPD$ by calculating the average log-likelihood across the heldout subsamples of our ARD matrix \citep{vehtari2017practical}  .

This process,
with the results shown in Table~\ref{tab:elpd},
identifies the true model as describing the 
data best, which is not immediately clear from the PPCs and degree 
estimates shown above, while agreeing with the subpopulation size estimate 
results. In real data, only the PPCs would be available for all models while cross-validation is possible only for models without the {\color{red}entry} 
-level parameters in some advanced ARD models.
{\color{red} While cross-validation can be used to perform model selection, other versions
of cross validation discussed above could also be useful for model assessment and comparison. For example, 
exact LOO-CV which removes all data about an ARD respondent (i.e., one individual's set of observations, as in version (b) above) could be used as a form of robustness check, similar to detecting influential observations or outliers in regression. 
Another potential approach for using LOO-CV would be to consider leaving out ego or alter groups, as in versions (c) and (d) above. 
This is especially promising since it summarizes model fit across the main entities of interest. 
However, this would need to be done carefully to ensure model stability 
and is related to the discussion of name selection in \citet{mccormick_how_2010}.  
In fact, retaining an opinion panel (or subset of columns corresponding to representative names) could provide a useful implementation of row-wise LOO-CV, as in (b), as it would preserve stability of model estimates and highlight model fit among the hard-to-reach subpopulations.}
This is strong evidence that general tools for model selection {\color{red}among} (all possible) ARD models are needed.

\begin{table}[ht!]
\centering
\begin{tabular}{|l|l l|}
\hline
\textbf{Model} & \textbf{Difference in $ELPD$}  & \textbf{$SE-ELPD$}\\
\hline
Latent Space & 0.0  & 0.0 \\
\hline
Overdispersed & $\num{-3.99e2}$ & $\num{2.96e+01}$ \\
\hline
Erdos-Renyi & $\num{-1.95e+04}$ & $\num{4.73e+02}$ \\
\hline
Varying Degree & $\num{-1.39e+10}$ & $\num{1.99e+07}$ \\
\hline
\end{tabular}
\caption{$K=10$ fold CV for the four models fitted to latent space data.
Differences in the estimates of the expected log-predictive density ($ELPD$) across the four models and the standard error of this quantity ($SE - ELPD)$ are given above. Here a value of $ELPD=0$,
indicates the best model, with negative values indicating worse
performance (compared to the model with $ELPD=0$. If the $ELPD$ is
larger than two times the estimated standard error, this can be interpreted as evidence
for a difference in predictive performance between the models.
There results indicate that the (true) Latent Space model has the best predictive performance for this data.}
\label{tab:elpd}
\end{table}

\subsection{Summary}
Fitting expressive latent variable
models to ARD is challenging in practice. 
Bayesian methods 
are popular in the literature as they provide a principled way to 
model such data, {\color{red}allowing flexible models to be fit.
While Bayesian methods allow for existing model checking 
strategies to be used,
these models also present practical implementation challenges.} 
In this work we have
{\color{red}carefully motivated and described}
several popular Bayesian ARD 
models and the steps required to fit them to data.
We provide both the code to generate {\color{red}synthetic data and fit all models
we have considered. These illustrate the important statistical and practical
considerations which need to be considered by users of these models.}

For the Latent Space data, the varying degree and overdispersed models provide 
accurate estimates of the degree distribution, 
while the latent space model provides the 
best 
overall predictive performance.  All three models accurately reproduce the 
distribution of possible respondent-to-subpopulation counts by subpopulation in the PPCs, 
but struggle to 
estimate the size of the unknown population correctly.
For the synthetic replication of the \citet{mccarty2001comparing} data, 
the varying degree model provides {\color{red}good} estimates of degree distribution. The 
true barrier effects model produces the best estimate of the unknown subpopulation size 
and best performance on the PPCs, although the overdispersed model performs only slightly 
worse in both cases.

These results highlight the challenges faced by users of these Bayesian
ARD models. Even in simulation settings where we compare our fitted
models against true parameters 
{\color{red}which are unavailable in practice,
our experiments highlight that model selection is a challenging task.}
With real data
posterior predictive checks and surrogate residuals can identify 
wrong models, {\color{red}they cannot be used to distinguish between two competing 
plausible models.}
While our cross-validation approach works well at identifying 
the correct model in our simulation setting, it cannot (currently) be 
applied to all ARD models in popular use. 
{\color{red}These results demonstrate that model selection is particularly
challenging for modern ARD models, and new statistical techniques may be required.
In real applications,}
selecting a final model will be guided by the primary goals of 
inference and the specific application under study.
For example, social network analysis may emphasize accurate estimation of 
personal network size, in which case the estimated posterior degree
distribution can be compared to previous estimates or external validation data,
between models with comparable performance in other model checking diagnostics.
Alternatively,
public health applications may focus on 
obtaining reasonable estimates of hidden population sizes (e.g., individuals 
with opioid use disorder or whom are unhoused), 
in which case a leave-one-out procedure could be used for known subpopulations
to compare models which pass PPCs and otherwise describe the data well.
Finally, policy-makers may care 
most about understanding social behaviors, and prefer to use a model which 
accounts for nonrandom mixing through overdispersion, perhaps at the expense 
of accurate parameter recovery,  provided the chosen model can be fit efficiently for the given data. 
{\color{red}In practice, selecting a final model will be guided by the primary goals of inference and the specific application under study.
Even in controlled settings, no single model consistently outperforms others across all evaluation criteria.
As a result, practitioners must rely on a combination of model checking diagnostics and comparison methods, tailored to the problem at hand, to select an appropriate model.}

\section{Discussion}
{\color{red}This work provides a modern overview of Bayesian ARD models and important considerations for using and choosing between such models. 
We provide a unified computational implementation of popular Bayesian ARD models, demonstrate 
practical challenges in model checking and comparison, and describe computational tradeoffs.
Accurate estimation of ARD parameters depends on
balancing model complexity, which may increase 
computational burden
with model performance,
with simpler formulations 
that may capture much of the underlying structure.}
We {\color{red}illustrate} several key challenges and steps in fitting
{\color{red}these}
models, with the use of simulated datasets. 
While there has been some previous work attempting to 
assess the fit of such Bayesian models, tools
{\color{red}that} can be readily applied to all models are
limited and many implementations require advanced 
knowledge
of Bayesian computation and software.
Our implementation of these models in Stan provides researchers with a coherent collection 
of common 
Bayesian models in a state-of-the-art Bayesian sampling software,
{\color{red}aiding reproducibility and accessibility}.  
Our 
incorporation of a
within-iteration rescaling procedure
eliminates a post-processing step that is typically 
required for the valid interpretation of parameter estimates.
This is one way to improve sampling algorithm run time and improves chain convergence diagnostics.

The datasets we examine in our model checking demonstration are synthetic data, simulated 
from an ARD model. While the {\color{red}simulation}
of the \citet{mccarty2001comparing} data uses 
parameter values from the true data,
our replicated data is only as complex as \citet{maltiel2015estimating}'s barrier effects 
model we use to simulate it.
As a result, neither dataset can replicate the complexity of human behaviors that are 
evident in real ARD observations.  For example, the heaping effects common to ARD will not 
show up in PPCs for our synthetic data. In practice, real ARD may result in poorer model 
fit than found for our synthetic data. However, the process of performing model checks 
and considering these checks in the context of applied inference goals as discussed above, 
translates directly to real world ARD.

None of the models we consider in our demonstration of Bayesian 
tools for model criticism in Section \ref{sec:checking} require external 
covariate information.  However, as discussed in Section \ref{sec:models}, 
many proposed models for ARD accommodate external covariate information, 
typically collected for the respondents, but occasionally for the 
subpopulations or, more rarely, a combination thereof, such as the level of 
respect covariates in \citet{paniotto2009estimating}'s ARD study. Some 
models explicitly depend on external covariate information to model 
nonrandom mixing.  While the posterior predictive checks and cross-
validation measures we examine can be performed for ARD models that include 
covariates, metrics to examine {\color{red}
the specific impact of covariates in ARD 
models 
and covariate selection 
are important considerations, and have recently been considered
in the literature \citet{laga2026evaluating}}.
A large number of Bayesian ARD models have now been proposed and
implemented for a wide range of important applications. However tools
to compare and evaluate several competing models 
{\color{red}remains an key open challenge.}
While
there has been valuable recent work in this direction,
how to do model selection (for example) for all potential ARD models
is unclear, and is a timely future direction to consider.
{\color{red}Taken together, these challenges highlight the need
for a unified framework for evaluating and comparing ARD models in practice. This
work puts existing research in the context of practical model checking workflows and highlights important directions for future research.}



 \begin{appendix}

\section{Prior distributions and hyperparameters in the Stan files}
We consider ARD entries, $y_{ik}$ of an $N \times K$ matrix, where $y_{ik}$ is the count of the number of members of subpopulation $k$ that respondent $i$ reports knowing.  In our Stan implementations, we generally implement flat diffuse prior distributions, which is standard in the literature; the exact specifications and hyperparameters are available in our .stan files on GitHub and restated here for clarity.  The rescaling procedures are implemented within our Stan files, but not represented here.

\paragraph{Erd\"{o}s-R\'{e}nyi model}

\begin{align*}
&\text{Priors:} & log(d) &\sim \text{Normal} (0, 25)\\
&& \beta_k & \overset{iid}{\sim} \text{Normal} (0,5) \\[.1in]
& \text{Likelihood:} & y_{ik} &\sim \text{Poisson}(d \beta_k)
\end{align*}

\paragraph{Varying degree model}

\begin{align*}
& \text{Priors:} & log(d_i) &\sim \text{Normal} (0, 25)\\
&& \beta_k & \overset{iid}{\sim} \text{Normal} (0,5) \\[.1in]
& \text{Likelihood:} & y_{ik} &\sim \text{Poisson}(d_i \beta_k)
\end{align*}

\paragraph{Overdispersed model}

\begin{align*}
& \text{Hyperpriors:} & \sigma_{\alpha} &\sim \text{Truncated Normal}(0,5) \\
&& \sigma_{\beta} & \propto I_{(0,\infty)} \\
&&\mu_{\beta} & \propto 1 \\[.1in]
& \text{Priors:} & \alpha_i &\overset{iid}{\sim} \text{Normal}(0,\sigma_{\alpha})\\
&&\beta_k &\overset{iid}{\sim} \text{Normal}(\mu_{\beta},\sigma_{\beta})\\
&& \frac{1}{\omega_k} & \propto I_{(0,1)}\\[.1in]
& \text{Likelihood:} & y_{ik} &\sim \text{Negative Binomial} \left( \frac{e^{\alpha_i +\beta_i}}{\omega_k-1}, \frac{1}{\omega_k-1} \right)
\end{align*}

\paragraph{Latent Space model}
\begin{align*}
& \text{Hyperpriors:} & \mu_{\alpha} & \propto 1\\
&& \sigma_{\alpha} &\sim \text{Normal}(0,5) \\
&&\mu_{\beta} & \propto 1 \\
&& \sigma_{\beta} & \propto \text{Normal}(0,5) \\[.1in]
& \text{Priors:} & z_i & \overset{iid}{\sim} \text{von-Mises Fisher}(\mu = (0, 0, 1), \kappa = 0)\\
&& \nu_k & \overset{iid}{\sim} \text{von-Mises Fisher}(\mu = (1, 0, 0),\kappa=0)\\
&& \zeta &\sim \text{Gamma}(2,1)\\
&& \eta_k & \overset{iid}{\sim} \text{Gamma}(2,1) \\
&& \alpha_i &\overset{iid}{\sim} \text{Normal}(\mu_{\alpha},\sigma_{\alpha})\\
&&\beta_k &\overset{iid}{\sim} \text{Normal}(\mu_{\beta},\sigma_{\beta})\\[.1in]
& \text{Likelihood:} & y_{ik} &\sim \text{Poisson} \left( e^{\alpha_i + \beta_k} \kappa(\zeta,\eta_k,\theta_{(z_i,\nu_k)}) \right)
\end{align*}

 \end{appendix}

\printbibliography

\end{document}